\documentclass[a4paper,11pt]{article}
\usepackage{jheppub} 

\makeatletter
\def\@fpheader{}
\makeatother

\usepackage{units}
\usepackage{amsmath}
\usepackage{graphicx}
\usepackage{dcolumn}
\usepackage{pbox}
\usepackage{amssymb}
\usepackage{epsfig}
\usepackage{slashed}
\usepackage{amssymb}
\usepackage{ mathrsfs }
\usepackage[usenames,dvipsnames]{xcolor}

\usepackage{multirow}
\usepackage{lineno}
\usepackage{autobreak}

\usepackage{booktabs}

\usepackage{comment}

\usepackage{subcaption}
\usepackage{graphicx}

\usepackage{bbold}

\definecolor{MyLightBlue}{rgb}{0.22,0.51,0.9}
\definecolor{BrickRed}{rgb}{0.8, 0.25, 0.33}
\RequirePackage{hyperref}
\hypersetup{colorlinks, citecolor=blue,linkcolor=BrickRed, urlcolor=MyLightBlue}

\usepackage{bbding}
\usepackage{pifont}
\usepackage{wasysym}
\usepackage{amssymb}
\usepackage{tabu}

\usepackage{autobreak}

\title{\Large Accidental Peccei-Quinn Symmetry From Gauged U(1)\\[0.03in] and a High Quality Axion}

\author[a]{\bf K.S. Babu,}
\emailAdd{babu@okstate.edu}
\affiliation[a]{Department of Physics, Oklahoma State University, Stillwater, OK 74078, USA}

\author[b]{\bf Bhaskar Dutta,}
\emailAdd{dutta@tamu.edu}
\affiliation[a]{Department of Physics and Astronomy, Mitchell Institute of Fundamental Physics and Astronomy, Texas A\&M University, College Station, TX  77843, USA}

\author[c]{\bf Rabindra N. Mohapatra}
\emailAdd{rmohapat@umd.edu}
\affiliation[b]{Maryland Center for Fundamental Physics \& Department of Physics, University of Maryland, College Park, Maryland 20742, USA}

\abstract{We construct explicit models that solve the axion quality problem originating from quantum gravitational effects. The general strategy we employ is to supplement the Standard Model and its grand unified extensions by an anomaly-free axial $U(1)_a$ symmetry that is gauged.  We show that for several choices of the gauge quantum numbers of the fermions, this setup leads to an accidental $U(1)$ symmetry with a QCD anomaly which can be identified as the  Peccei-Quinn (PQ) symmetry that solves the strong CP problem.  The $U(1)_a$ gauge symmetry controls the amount of explicit PQ symmetry violation induced by quantum gravity, resulting in a high quality axion.  We present two classes of models employing this strategy. In the first class  (models I and II), the axial $U(1)_a$ gauge symmetry acts on vector-like quarks leading to an accidental KSVZ-type axion.  The second class (model III) is based on $SO(10)$ grand unified theory extended by a gauged $U(1)_a$ symmetry that leads to a hybrid KSVZ--DFSZ type axion. The couplings of the axion  to the electron and the nucleon are found to be distinct in this class of hybrid models from those in the KSVZ and DFSZ models, when the axion is identified as the dark matter of the universe, which can be used to test these models. Interestingly, all models presented here have domain wall number of one,  which is free of cosmological problems that typically arise in axion models.}

\begin{document}
\maketitle
\flushbottom

\section{Introduction}

Despite its many successes, the Standard Model (SM) of particle physics is plagued by an uncontrolled amount of CP violation in the strong interaction sector arising from non-perturbative QCD effects~\cite{Belavin:1975fg, tHooft:1976snw,Callan:1976je,Jackiw:1976pf}. This CP violation can be characterized by a new interaction Lagrangian given by $(g_s^2/32\pi^2) \,\overline{\theta} \,G_{\mu\nu}^a\tilde{G}^{a,\mu \nu}$, where $g_s$ is the strong coupling constant, $G_{\mu\nu}^a$ is the gluon field strength tensor, $\tilde{G}^{a,\mu\nu}$ is its dual and $\overline{\theta} $ is a dimensionless coupling. In presence of this interaction, which is also parity-violating, the neutron would acquire an electric dipole moment (EDM) $d_n, $ of order $d_n \sim (\overline{\theta} \times 10^{-16}$) e-cm~\cite{Crewther:1979pi}. The current experimental limit on neutron EDM, $d_n \leq 1.1 \times 10^{-26}$ e-cm \cite{Abel:2020pzs}, implies that $\overline{\theta}  \leq 10^{-10}$. A lack of understanding of the extreme smallness of this dimensionless parameter is known as the strong CP problem. Invoking CP as an approximate symmetry to explain its smallness does not appear to be natural, since the analogous CP violating parameter in the weak interaction sector, the CKM phase $\delta$, is of order unity.  

The most widely discussed solution to the strong CP problem is the Peccei-Quinn mechanism~\cite{Peccei:1977hh}, where one postulates the existence of a light pseudoscalar boson, the axion, $a(x)$, which arises as the pseudo-Nambu-Goldstone boson associated with the spontaneous breaking of a global $U(1)_{\rm PQ}$ symmetry \cite{Weinberg:1977ma,Wilczek:1977pj}. This global $U(1)_{\rm PQ}$ symmetry is also broken explicitly by a QCD anomaly, which induces a coupling of the axion to the gluon fields given by $(g_s^2/32\pi^2) \,\left(\frac{a(x)}{f_a}\right) \,G_{\mu\nu}^a\tilde{G}^{a,\mu \nu}$,  where $f_a$, the axion decay constant, is the vacuum expectation value (VEV) of a scalar field that spontaneously breaks the PQ symmetry. The axion develops a potential, also through the QCD anomaly, which
appears as an even function of the combination $(\overline{\theta} +\frac{a}{f_a})$. Minimizing this potential leads to the condition $\frac{a}{f_a}+\overline{\theta} =0$, which dynamically relaxes the $\overline{\theta}$ parameter to zero and thus solves the strong CP problem.  

A prototypical axion model has the following effective Lagrangian:
\begin{equation}
{\cal L}^{\rm eff} = \frac{g_s^2}{32 \pi^2}\frac{a(x)} {f_a} G_{\mu \nu}^a \tilde{G}^{a,\mu \nu} - \lambda(|\Phi|^2-f_a^2/2)^2 - \Lambda^4 \cos \left(\frac{Na(x)}{f_a}\right),
\label{eq:axionpot1}
\end{equation}
where $\Phi$ is a complex scalar field acquiring a vacuum expectation value $f_a$, and parametrized as $\Phi = (\rho + f_a)/\sqrt{2}\,e^{ia(x)/f_a}$, with $\Lambda \sim 200$ MeV being the QCD confinement scale. The last term in Eq. (\ref{eq:axionpot1}) arises from non-perturbative QCD effects which break the $U(1)_{\rm PQ}$ symmetry explicitly.   Here $N$ is an integer greater than or equal to 1 determined by the QCD anomaly coefficient. The explicit QCD breaking term of $U(1)_{\rm PQ}$ still preserves a $Z_N$ symmetry, under which $a(x)\to a(x)+ 2\pi f_a m/N$ ($m=0, 1,...,N-1)$. This unbroken discrete symmetry gives rise to domain walls, which can modify the evolution of the universe drastically~\cite{Sikivie:1982qv}, an issue which we shall return to later. Minimizing the potential for the axion field given in Eq. (\ref{eq:axionpot1}) leads to solution $a/f_a = 0$. In writing Eq. (\ref{eq:axionpot1}) we have made use of a shift symmetry, $a(x) \rightarrow a(x) + f_a \,\overline{\theta}$, present in these models, and absorbed the $\overline{\theta}$ term into the axion field.

Being a pseudo-Goldstone boson, the coupling of the axion to fermions are derivative in nature, given as
\begin{equation}
{\cal L}_{a \bar{f} f} =\frac{\partial_\mu a}{2 f_a}\left(C_{af}\overline{f}_i \gamma^\mu \gamma_5 f_i\right)~\equiv~ g_{af}\frac{\partial_\mu a}{2 m_f}\left(\overline{f}_i \gamma^\mu \gamma_5 f_i\right)~.
\label{eq:Cf}
\end{equation}
Here $C_{af}$ (or equivalently $g_{af}$) are model-dependent coefficients which are crucial for experimental detection of the axion. 
Since the axion is an ultralight particle, if $f_a$ is of order of the weak scale, it would have been produced in many laboratory experiments, such as kaon decay $K^+ \rightarrow \pi^+ + a$, as well as in astrophysical settings such as cooling of stars, leading to visible signals. The absence of any such signal rules out weak scale axion models~\cite{Weinberg:1977ma, Wilczek:1977pj}.  
This has led to the construction of two classes of {\it invisible axion} models: (i) the Dine-Fischler-Srednicki-Zhinitsky (DFSZ) model~\cite{Dine:1981rt,Zhitnitsky:1980tq}, where the SM fermions are charged under a global $U(1)_{\rm PQ}$ symmetry which has a QCD anomaly, and where fermion mass generation is realized through a second Higgs doublet, and (ii) the Kim-Shifman-Vainshtein-Zakharov (KSVZ) model~\cite{Kim:1979if,Shifman:1979if} which introduces new vector-like quarks carrying $U(1)_{\rm PQ}$ charges, with the SM fermions being neutral under the PQ symmetry.  In both cases the $U(1)_{\rm PQ}$ breaking scale is in the range $f_a = (10^{9}-10^{12})$ GeV, making the interactions of axion with matter very feeble.  These limits on $f_a$ are primarily derived from stellar cooling, supernova 1987A observations~\cite{Raffelt:1987yt,Turner:1987by}, as well as the cosmological mass density of the universe~\cite{Preskill:1982cy, Abbott:1982af, Dine:1982ah} (for recent reviews see Ref.~\cite{DiLuzio:2020wdo,GrillidiCortona:2015jxo,Marsh:2015xka}).


The high scale of invisible axion models comes with a price. These models suffer from the so-called axion quality problem caused by quantum gravitational effects~\cite{Kamionkowski:1992mf,Holman:1992us,Barr:1992qq,Ghigna:1992iv}. All global symmetries of nature are believed to be broken by non-perturbative gravity effects such as black holes or worm holes. Such breakings should also apply to the $U(1)_{\rm PQ}$ symmetry. These breaking effects may be parameterized by higher dimensional operators suppressed by appropriate powers of the Planck scale, $M_{\rm Pl}$.  If an operator of the type $|\Phi|^4 \Phi/M_{\rm Pl}$, induced by quantum gravity, is added to the scalar potential of Eq. (\ref{eq:axionpot1}), the $\overline{\theta} = 0$ solution would be destabilized, unless the coefficient of this operator is less than about $10^{-50}$. The required tuning of parameters here is much stronger than the one needed for the original $\overline{\theta}$. 
This in essence is the axion quality problem~\cite{Kamionkowski:1992mf,Holman:1992us,Barr:1992qq,Ghigna:1992iv}.

There have been various attempts in the literature to address the axion quality problem. One idea that has been suggested is to realize $U(1)_{\rm PQ}$ as an accidental symmetry in presence of a gauged $U(1)$ symmetry~\cite{Barr:1992qq,Qiu:2023los,Babu:2024udi}.  Such an accidental axion may also arise from non-Abelian gauge symmetries~\cite{DiLuzio:2020qio,Ardu:2020qmo}. A second mechanism assumes that the axion is a composite particle, in which case the quantum gravity-induced operators may have high enough power suppression factors~\cite{Randall:1992ut,Lillard:2018fdt,Gaillard:2018xgk,Vecchi:2021shj,Lee:2018yak,Cox:2023dou, Cox:2021lii,Nakai:2021nyf, Contino:2021ayn,Podo:2022gyj}. Discrete gauge symmetries~\cite{Babu:2002ic} have been proposed that act as accidental $U(1)_{\rm PQ}$ up to very high orders in non-renormalizable operators. Mirror universe models~\cite{Berezhiani:2000gh, Hook:2019qoh}, multiple replication of the SM sector~\cite{Hook:2018jle, Banerjee:2022wzk},  and extra dimensional~\cite{Choi:2003wr, Reece:2024wrn, Craig:2024dnl} and string theoretic constructions~\cite{Svrcek:2006yi} have also been proposed to address the axion quality problem. Extension of the color gauge symmetry from $SU(3)$ to $SU(N)$ has been proposed to generate new contributions to the axion mass dynamically~\cite{Gherghetta:2016fhp}, which can also help address the axion quality problem.

The purpose of this paper is to present a class of models that provides a high-quality axion. 
The method we develop here is based on a gauged $U(1)_a$ extension of the SM as well as a unified $SO(10)$ theory which naturally leads to an accidental $U(1)_{\rm PQ}$ symmetry by virtue of the particle spectrum and the $U(1)_a$ charge assignment. This method is similar to the one suggested in Ref.~\cite{Barr:1992qq}, but with a more economical particle spectrum, and an extension to grand unification. We present two classes of models. In the first class (models I and II), SM is extended with a gauged $U(1)_a$ symmetry acting on a set of vector-like quarks that results in an accidental $U(1)_{\rm PQ}$ and a KSVZ-type axion.  The second class of models is based on unified $SO(10)$ symmetry appended by a gauged $U(1)_a$ that results in a hybrid DFSZ-KSVZ axion model (model III).   A common feature of these models is that they all have two disjoint sectors which are connected only by higher-dimensional gravity-induced operators. At the renormalizable level these models have a $U(1)_a \times U(1)_{\rm PQ}$ symmetry, with the $U(1)_{\rm PQ}$ being anomalous with respect to QCD.  As long as the two sectors are sufficiently separated, measured in terms of the dimensionality of the gravity-induced operators that generate interlocking couplings, a high quality axion can be realized. The gauged $U(1)_a$ symmetry protects quantum gravity corrections to $\overline{\theta}$ to sufficient degree in these models, thus ensuring a high quality axion.

The paper is organized as follows. In Sec.~\ref{sec:sec2} 
we discuss quantitatively the axion quality problem and provide an overview of the mechanism that we adopt to solve it via gauged $U(1)_a$ symmetry. In Sec.~\ref{sec:Model1} we present our first model (model I) that results in an accidental KSVZ-type axion. Four vector-like quarks with axial $U(1)_a$ charges are employed in model I.  Here we also discuss the phenomenology and the cosmology of the model as well as the fermion mass generation mechanism.  Sec.~\ref{sec:Model2} generalizes model I into a family of models (model II), which introduces $m+1$ vector-like quarks where $m \geq 4$. In Sec.~\ref{sec:Model3} we present a unified $SO(10) \times U(1)_a$ model that leads to a high-quality hybrid DFSZ-KSVZ axion model which we proposed in Ref.~\cite{Babu:2024udi}. In Sec.~\ref{sec:Model3-pheno} we provide further details of the model, including loop-induced corrections to the $\overline{\theta}$ parameter, calculation of the domain wall number, and phenomenological issues and dark matter parameter space of the hybrid axion.
In Appendix~\ref{sec:Appendix-A}, we briefly review the Barr-Seckel model of high quality axion with which our models I and II may be compared. Here we also provide a UV completion of this class of models and discuss ways to address the cosmological domain wall issues.

\section{The axion quality problem and a solution from gauged \boldmath{$U(1)_a$}}

\label{sec:sec2}

In this section, we provide a quantitative analysis of the axion quality problem and give an overview of our method to address it with a gauged $U(1)_a$ symmetry.

\subsection{Axion quality problem}

The explicit breaking of the global $U(1)_{\rm PQ}$ symmetry by the QCD anomaly induces a coupling of the axion field to the gluon fields with the effective Lagrangian given as 
$(g_s^2/32\pi^2) \,\left( \frac{a(x)}{f_a}+ \overline{\theta}\right)\,G_{\mu\nu}^a\tilde{G}^{a,\mu \nu}$. This coupling, in turn, would induce a potential for the axion field, which has been computed in chiral perturbation theory to be~\cite{Weinberg:1977ma,DiVecchia:1980yfw}:


\begin{equation}
V(a) \simeq  -m_\pi^2 f_\pi^2 \sqrt{1-\frac{4 m_u m_d}{(m_u+m_d)^2} \sin^2 \left(\frac{a}{2 f_a} + \frac{\overline{\theta}}{2}\right)}~.
\label{eq:axionpot2}
\end{equation}
Minimizing this potential would lead to the desired condition
\begin{equation}
    \sin\left(\frac{a}{f_a}+\overline{\theta}\right) = 0~.
    \label{eq:mini}
\end{equation}
(Here the $\overline{\theta}$ term is not absorbed into the axion field by a shift transformation $a \rightarrow a + f_a \overline{\theta}$, unlike in Eq. (\ref{eq:axionpot1}).) 
Once  we add to this  potential higher dimensional operators induced by quantum gravity that violate the PQ symmetry, $\left(\frac{a}{f_a}+\overline{\theta}\right)$ would shift away from zero, destabilizing the axion solution to the strong CP problem. 

Consider, for example, the addition of a $d=5$ term to the Higgs potential of Eq. (\ref{eq:axionpot1}) arising from quantum gravity:
\begin{equation}
V_{\rm gravity} = \frac{\kappa}{2! \,3!\,M_{\rm Pl}} |\Phi|^4 (e^{i \delta} \Phi + h.c.)
\label{eq:vgravity}\end{equation}
Here, $\kappa$ and $\delta$ are real parameters, which are presumably of order unity. 
The factors $2!\,3!$ in the denominator of Eq.~(\ref{eq:vgravity}) are symmetry factors for identical $(\Phi^*)^2$ and $(\Phi)^3$ fields, which we shall adopt for quantum gravity corrections. (We also quote results when such symmetry factors are assumed to be not present.)  Writing $\Phi = (\rho + f_a)/\sqrt{2}\, e^{i a(x)/f_a}$, Eq. (\ref{eq:vgravity}) leads to a correction to the axion potential given as
\begin{equation}
V_{\rm gravity}^{(a)} \simeq  \frac{\kappa}{6 M_{\rm Pl}}\frac{f_a^5}{2^{5/2}}\cos\left(\frac{a}{f_a}+\delta \right)~.
\label{veqn2}\end{equation}
Minimizing the full potential of Eq.~(\ref{eq:axionpot2}) and Eq.~(\ref{veqn2}) would shift the vacuum so that $\left(\frac{a}{f_a} +\overline{\theta}\right)$ now becomes (assuming this shift arising from $V_{\rm gravity}^{(a)}$ is small)
\begin{equation}
\left(\frac{a}{f_a} +\overline{\theta}\right)\simeq \frac{\kappa \sin\delta}{24 \sqrt{2}}\frac{f_a^5}{m_\pi^2 f_\pi^2 M_{\rm Pl}} \frac{(m_u+m_d)^2}{m_u \,m_d}~.
\end{equation}
Using $M_{\rm Pl} = 1.22 \times 10^{19}$ GeV, $f_\pi = 93$ MeV, $m_\pi = 140$ MeV, and $m_u/m_d = 0.56$, and requiring that $\overline{\theta} \leq 10^{-10}$ leads to the constraint
\begin{eqnarray}
|\kappa \sin\delta| \leq  1.6 \times \left(10^{-39},\,10^{-44},\, 10^{-54}\right),~~{\rm for}~~f_a =  \left(10^9,\, 10^{10},\,10^{12}\right)~{\rm GeV} ~.
\end{eqnarray}
The required fine-tuning is even more severe than the one needed for the original strong CP parameter $\overline{\theta}$, causing the axion quality problem~\cite{Kamionkowski:1992mf,Holman:1992us,Barr:1992qq,Ghigna:1992iv}.
Note that in the original Weinberg-Wilczek axion model (when supplemented by a singlet scalar $\Phi$, which also acquires an electroweak scale VEV), the required tuning is very mild, with $|\kappa \sin\delta| \leq 10^{-3}$ for $f_a = 10^2$ GeV. It is the phenomenological requirement of making the axion invisible, by choosing $f_a =(10^9-10^{12})$ GeV, that results in the quality problem.

\subsection{Mechanism for solving the axion quality problem with a gauged \boldmath{$U(1)_a$}} 

The models we present have a gauged $U(1)_a$ symmetry which acts axially on fermion fields.  That is, the $U(1)_a$ charges of left-chiral and right-chiral fermions are equal and opposite.  In addition, the fermions and scalar fields form two separate sectors at the renormalizable level, leading to an accidental $U(1)_p \times U(1)_q$ symmetry.  Each of these accidental $U(1)$s has a QCD anomaly, but one combination must have vanishing QCD anomaly, corresponding to the gauged $U(1)_a$ symmetry.  The full symmetry may then be taken as $U(1)_a \times U(1)_{\rm PQ}$, with the second factor leading to an accidental axion.

To see more explicitly the origin of the axion field, consider the breaking of the $U(1)_a$ gauge symmetry by two SM singlet scalar fields $S$ and $T$ with $U(1)_a$ charges of $q_S$ and $q_T$.  If the charges $q_{S,T}$ are appropriately chosen, the potential for the $S$ and $T$ fields at the renormalizable level may be functions of the polynomials $S^* S$ and  $T^* T$ with no interlocking terms that are nontrivial. The quark fields also belong to two separate sectors, with one set of quarks acquiring their masses from the VEV of $S$, and the other from $T$. The renormalizable theory will then have a global symmetry $U(1)_{S}\times U(1)_{T}$, operating on the $S$ and $T$ sectors separately. 
The axion field arises as the orthogonal combination to the Nambu-Goldsonte field that is eaten up by the $U(1)_a$ gauge boson.  To see this, let us parametrize the $S$ and $T$ fields, which are assumed to have negative squared masses, as
\begin{equation}
T = \frac{(\rho_T + f_T)}{\sqrt{2}} e^{i\eta_T/f_T},~~~S = \frac{(\rho_S + f_S)}{\sqrt{2}} e^{i\eta_S/f_S}~.
\label{eq:param1}
\end{equation}
Here $(\rho_S,\,\rho_T)$ are the radial modes, which do not play any role in the axion identification, and may be set to zero for this purpose. $f_S$ and $f_T$ are the VEVs of the respective fields $S$ and $T$.  The Goldstone mode and the orthogonal axion mode are then found to be
\begin{eqnarray}
G = \frac{(q_S f_S \eta_S + q_T f_T \eta_T)}{\sqrt{q_S^2 f_S^2 + q_T^2 f_T^2}},~~a = \frac{(q_T f_T \eta_S - q_S f_S \eta_T)}{\sqrt{q_S^2 f_S^2 + q_T^2 f_T^2}}~.
\end{eqnarray}
Owing to the accidental $U(1)_{\rm PQ}$ symmetry, the axion field receives zero mass from the scalar potential, although the QCD anomaly would induce a potential for it as given in Eq. (\ref{eq:axionpot2}), and from it a small axion mass given by
\begin{equation}
m_a^2 \simeq \frac{f_\pi^2 m_\pi^2}{f_a^2} \frac{m_u m_d}{(m_u+m_d)^2}~.
\label{eq:axionmass}
\end{equation}

Higher-dimensional gravity-induced operators would generate interlocking couplings involving $T$ and $S$ in non-trivial ways.  If the charges $q_S$ and $q_T$ are such that these operators have high enough dimension, the axion quality can be preserved, as we show in the three models discussed in subsequent sections. As an example, consider the charges $(q_S,\,q_T)$ to be co-prime integers, in which case the lowest order interlocking operator induced by gravity would be $(S)^{q_T} (T^*)^{q_S}/M_{\rm Pl}^{q_S+q_T-4}$.  If in any given model, $(q_S + q_T) \geq 10$ can be satisfied, a high-quality axion will be realized~\cite{Kamionkowski:1992mf,Barr:1992qq}.

\section{ Model I: Realizing high-quality KSVZ--type axion} 

\label{sec:Model1}

The first model we present is based on the gauge group $G_{SM}\times U(1)_a$, where $G_{SM}$ is the standard model gauge group.  All SM fermions and the Higgs doublet are assumed to be neutral under $U(1)_a$. We add  four vector-like quarks (VLQs) which are color triplets and $SU(2)_L$ singlets carrying hypercharge $y$.\footnote{We find that the minimum number of VLQs needed to realize a high-quality axion in this framework is 4.  With only 3 such VLQs, having axial $U(1)_a$ charges of $\pm(1,1,-2)$, loop corrections to the shift in $\overline{\theta}$ destabilize the axion solution.} Their charge assignment is given  in Table~\ref{tab:tab1}. It is easy to see that all gauge anomalies cancel.  All linear anomalies in $U(1)_a$ vanish among the left-handed and the right-handed VLQs separately owing to the tracelessness property of the charges:
\begin{equation}
\sum_{i=1}^{4} q_i(Q_{L}) = 0,~~~~\sum_{i=1}^{4} q_i(Q_{R}) = 0~.
\end{equation}
This includes anomaly coefficients $A[SU(3)_c^2 \times U(1)_a]$, $A[SU(2)_L^2 \times U(1)_a]$, $A[U(1)_Y^2 \times U(1)_a]$ and $A[({\rm gravity})^2 \times U(1)_a]$, which all vanish among the left-handed fermions and separately among the right-handed fermions.  As for the anomaly coefficient $A[U(1)_Y \times U(1)_a^2]$, the axial assignment of the $U(1)_a$ charges, along with the universal hypercharge assignment $y$, implies that the left-handed and right-handed VLQs contribute an equal and opposite amount, leading to the cancellation of this anomaly as well. Finally, the cubic anomaly coefficient $A[(U(1)_a)^3]$ adds up to zero when the VLQ contributions ($-144$) and the right-handed singlet fermion contributions  $N_{Ri}$ ($+144$) are summed. The choice of $y=2/3$ or $-1/3$ has the advantage that the VLQs can mix with the SM quarks through higher dimensional operators, in which case there are no cosmological problems with a stable VLQ even with post-inflationary PQ symmetry breaking.

\begin{table}[t] 
   \label{tab:example}
   \small
   \centering
   \begin{tabular}{|c|c|c|}
   \hline
   \textbf{New fermion} & \textbf{\boldmath{$SU(3)_c \times SU(2)_L \times U(1)_Y$} } & \textbf{\boldmath{$U(1)_a$} charge $q_i$}  \\ 
   \hline
   $(Q_L)_{1,2,3,4}$& $(3,1,y)$ & ~$(1,1,1,-3)$ \\
   \hline
   $(Q_R)_{1,2,3,4}$ & $(3,1,y)$ & $-(1,1,1,-3)$ \\
   \hline
     $(N_R)_{1,2,3}$& $(1,1,0)$ & ~$(2,4,-6)$ \\
   \hline
   \textbf{New scalar} & \textbf{\boldmath{$SU(3)_c \times SU(2)_L \times U(1)_Y$} } & \textbf{\boldmath{$U(1)_a$} charge}\\
\hline
$T$& $(1,1,0)$ & $6$ \\
   \hline
   $S$ & $(1,1,0)$ & ~~~$2/n$ \\
   \hline\hline 
   \end{tabular}
   \caption{New fermions and scalars, along with their quantum numbers under the standard model as well as their respective  $U(1)_a$ charges $q_i$ in Model I. Here $n \geq 4$ is an integer.}
   \label{tab:tab1}
\end{table}

The model has two scalar fields denoted as $T$ and $S$ with $U(1)_a$ charges 6 and $2/n$ respectively.  Here $n$ is an integer greater than or equal to 4.  This choice is dictated by the requirement of high-quality axion in presence of gravity-induced higher dimensional operators, including loop corrections that modify the Higgs potential.  
The Higgs potential involving the scalar fields $S$ and $T$ is given by
\begin{eqnarray}
 V(T, S) = - \mu^2_T\, T^* T-\mu^2_S \,S^* S +\lambda_S (S^* S)^2 +\lambda_T (T^* T)^2
 +\lambda_{TS} (T^* T)(S^* S)~.
 \label{eq:potential}
\end{eqnarray}
As noted in Sec.~\ref{sec:sec2}, the potential of Eq. (\ref{eq:potential}) has no interlocking terms, at the renormalizable level, connecting the $T$ and $S$ fields nontrivially.

Bare mass terms for the vector-like quark fields and the singlet fermion fields $N_{iR}$ in the model are not allowed owing to the chiral nature of their $U(1)_a$ charges. Their masses are generated through 
 Yukawa couplings given by
 \begin{eqnarray}
 -{\cal L}_{\rm Yuk} &=&Y^{(Q)}_{44} \,\overline{Q}_{4L}\,Q_{4R}\,T^*+\sum_{a,b=1}^3 Y^{(Q)}_{ab} \,\overline{Q}_{aL}Q_{bR}\frac{S^n}{n!\,M^{ n-1}_*} \nonumber \\
 &+&Y^{(N)}_{12} N_{1R} N_{2R} T^* + \frac{Y^{(N)}_{33}}{2} N_{3R} N_{3R} \frac{T^2}{2!\,M_*} + h.c.
 \label{eq:Yuk1}
 \end{eqnarray}
Here $M_*$ is a new scale, which should not be much larger than the PQ breaking scale, so that the VLQs acquire masses above a TeV to be compatible with LHC search limits. For example, with $n=4$, the choice of $f_S = 5 \times 10^{10}$ GeV would set an upper limit of $M_* \leq 6 \times 10^{12}$ GeV, so that the VLQ masses are greater than 1.2 TeV, where we have assumed that the Yukawa coefficients $Y^{(Q)}_{ab}$ are of order unity.

Note that this model has an accidental $U(1)\times U(1)$ symmetry.  The $U(1)$ charges of various fields may be assigned as follows. Under the first $U(1)$, $(Q_{4L},\,Q_{4R})$ have charges $(-3,\,+3)$, $(N_{1R},\,N_{2R},\,N_{3R}) = (2,\,4,\,-6)$ and $T=+6$ while all other fields have zero charge.  Under the second $U(1)$, $(Q_{aL},\,Q_{aR})=(1,\,-1)$ for $a=1,2,3$, and $S=2/n$ with other fields being neutral. Thus, $(Q_{4L},\,Q_{4R}, N_{(1,2,3)R},\,T)$ form one sector of the theory while $(Q_{aL},\,Q_{aR},\,S)$ form a second decoupled sector. Both of these $U(1)$s have QCD anomalies, but one combination of the two is the gauged $U(1)_a$, which has no anomalies. An orthogonal combination to $U(1)_a$ is a global symmetry which has a QCD anomaly and serves as the $U(1)_{\rm PQ}$ symmetry.  The gravity-induced couplings, to be discussed in Sec.~\ref{sec:sec3.2}, would only respect the gauged $U(1)_a$ symmetry and would violate the global $U(1)_{\rm PQ}$ symmetry. 

\subsection { Calculation of $a\, G \tilde{G}$ coupling in Model I}

With negative squared masses for the $T$ and $S$ fields, both will acquire real VEVs, denoted as $f_T$ and $f_S$, as parametrized in Eq. (\ref{eq:param1}). The masses of the vector-like fermions and the singlet fermions are given by
\begin{eqnarray}
M_{Q_4} &=& \frac{Y^{(Q)}_{44} f_T}{\sqrt{2}},~~~~~~~~~~~~~~~ M_{Q_a} = \frac{Y^{(Q)}_a f_S^n}{2^{n/2} \,n!\, M_*^{n-1}}~~(a=1,2,3) \nonumber \\
M_{N_1} &=& M_{N_2} = \frac{Y^{(N)}_{12} f_T}{\sqrt{2}},~~~~~M_{N_3} = \frac{Y^{(N)}_{33} f_T^2}{4 M_*}~.
\end{eqnarray}
Here $|Y_a^{(Q)}|^2$ are the eigenvalues of the $3 \times 3$ matrix $Y^{{(Q)}^\dagger} Y^{(Q)}$ defined in Eq. (\ref{eq:Yuk1}).  With $M_*$ assumed to be not too much larger than the VEVs $(f_T,\,f_S)$, all vector-like fermions acquire masses of order $(f_T,\,f_S)$.

Using the exponential parametrization for $(T,\,S)$ fields as in Eq. (\ref{eq:param1}), we can identify the would-be Goldstone field $G$ eaten up by the $U(1)_a$ gauge boson and the orthogonal axion field $a$ as:
\begin{eqnarray}
G = \frac{\left(\frac{2}{n}f_S \eta_S + 6 f_T \eta_T  \right)}{\sqrt{(6 f_T)^2+ (\frac{2}{n}f_S)^2}},~~~a = \frac{\left(6 f_T \eta_S -\frac{2}{n}f_S \eta_T  \right)}{\sqrt{(6 f_T)^2+ (\frac{2}{n}f_S)^2}}~.
\label{eq:axionmod1}
\end{eqnarray}
The Lagrangian involving the quarks and the axion field can now be readily obtained.  We find it to be
\begin{equation}
    -{\cal L}_{\rm Yuk}(a) = i a \left[ \overline{Q}_4 \gamma_5 Q_4 \frac{f_S}{f_T} \frac{M_{Q_4}}{ \sqrt{9 f_T^2 n^2 + f_S^2}} + \sum_{a=1}^{3} \overline{Q}_{a} \gamma_5 Q_{a}\frac{3 n^2 f_T}{f_S} \frac{M_{Q_{a}}}{\sqrt{9 f_T^2 n^2 + f_S^2}}
    \right]~.
\end{equation}
Summing over the four quarks in computing $a\,G \tilde{G}$ coefficient, one obtains
\begin{equation}
\frac{1}{f_a} = \frac{f_S}{f_T}\frac{1}{\sqrt{9 f_T^2 n^2 + f_S^2}} + (3 \times 3 n^2) \frac{f_T}{f_S} \frac{1}{\sqrt{9 f_T^2 n^2 + f_S^2}}~,
\end{equation}
which simplifies to
\begin{equation}
f_a = \frac{f_T f_S}{\sqrt{9 f_T^2 n^2 + f_S^2}}~.
\label{eq:famod1}
\end{equation}
It appears in the axion-gluon interaction Lagrangian defined as
\begin{equation}
    {\cal L}_{aGG} = \frac{g_s^2}{32 \pi^2} \frac{a}{f_a}G_{\mu \nu}^a \tilde{G}^{a,\mu \nu}~.
\end{equation}

\subsection{ Gravity effects and axion quality in Model I }
\label{sec:sec3.2}

Since gravity effects only respect the gauge symmetry, and not the global $U(1)_{\rm PQ}$ symmetry, one should ensure that these effects, which are parametrized as higher dimensional operators with appropriate powers of inverse Planck mass, do not destabilize the PQ solution to the strong CP problem.  The separation of the fields into two disjoint sectors,  $(Q_{4L},\,Q_{4R}, N_{(1,2,3)R},\,T)$ and $(Q_{aL},\,Q_{aR},\,S)$ is spoiled by gravity-induced corrections.
In this section, we investigate this issue for model I. There are several sources of $U(1)_{\rm PQ}$ symmetry violation.  The first source we analyze is the contribution to the scalar potential of the model, with the leading term given by

\begin{equation}
V_{\rm gravity} = \frac{\kappa e^{i \delta} S^{3n} T^*}{(3n)!\, M_{\rm Pl}^{3n-3}} + h.c.
\label{eq:pot3}
\end{equation}
The potential involving the axion field resulting from here is
\begin{equation}
V_{\rm gravity}^{(a)} = \frac{\kappa}{(3n)!\, 2^{(3n-1)/2}} \frac{f_S^{3n} f_T}{M_{\rm Pl}^{3n-3}} \cos\left(\frac{a}{f_a}+ \delta  \right)
\label{eq:Vgravitya}
\end{equation}
Minimizing this potential, along with the QCD-induced axion potential given in Eq. (\ref{eq:axionpot2}), will lead to a shifted value of $\overline{\theta}$, which, in the approximation that the shift is small, is given as
\begin{equation}
\overline{\theta} \simeq \frac{\kappa \sin\delta}{(3n)!\,2^{(3n-1)/2}} \frac{f_S^{3n} f_T}{M_{\rm Pl}^{3n-3}m_\pi^2 f_\pi^2} \frac{(m_u+m_d)^2}{m_um_d}~.
\label{eq:shiftmod1}
\end{equation}
One can also express the $\overline{\theta}$ induced by gravity in terms of the induced axion mass as
\begin{equation}
\overline{\theta}\simeq  \sin\delta\, \frac{m^2_{a,{\rm gravity}}}{m_a^2},
\label{eq:gravmass}
\end{equation}
where $m_{a,{\rm gravity}}$ is the axion mass arising from $V_{\rm gravity}^{(a)}$ of Eq. (\ref{eq:Vgravitya}). Thus, for a high quality axion, the gravity-induced axion mass should be $\leq 10^{-5} \times m_a$, with $m_a$ arising from QCD effects as given as in Eq. (\ref{eq:axionmass}). This is a generic feature which applies to all the subsequent models that we discuss.  

Choosing $f_S = f_T = f_a\sqrt{1+9 n^2}$, we find, with $\kappa \sin\delta = 1$,
\begin{equation}
  |\overline{\theta}| \simeq
    \begin{cases}
      2.7 \times 10^{-25} ~~ (n=4,\,f_a = 5 \times 10^{10}~{\rm GeV})\\
      9.6 \times 10^{-29} ~~(n=5, \, f_a = 10^{12}~{\rm GeV})~.
    \end{cases}       
\end{equation}
This shows that the axion is of high quality, including the gravity-induced higher dimensional operator of Eq. (\ref{eq:pot3}). If $n=3$ is chosen, the axion will not be of high quality, since  $|\overline{\theta}| \simeq 0.48$ will result for $f_a = 5 \times 10^{10}$ GeV, a value preferred by axion being the entire dark matter component of the universe. The choice of $n=3$ is also disfavored by loop-induced contributions to $\overline{\theta}$, which we now discuss.

\begin{figure}[t!]
		\centering
		\includegraphics[width=0.9\textwidth]{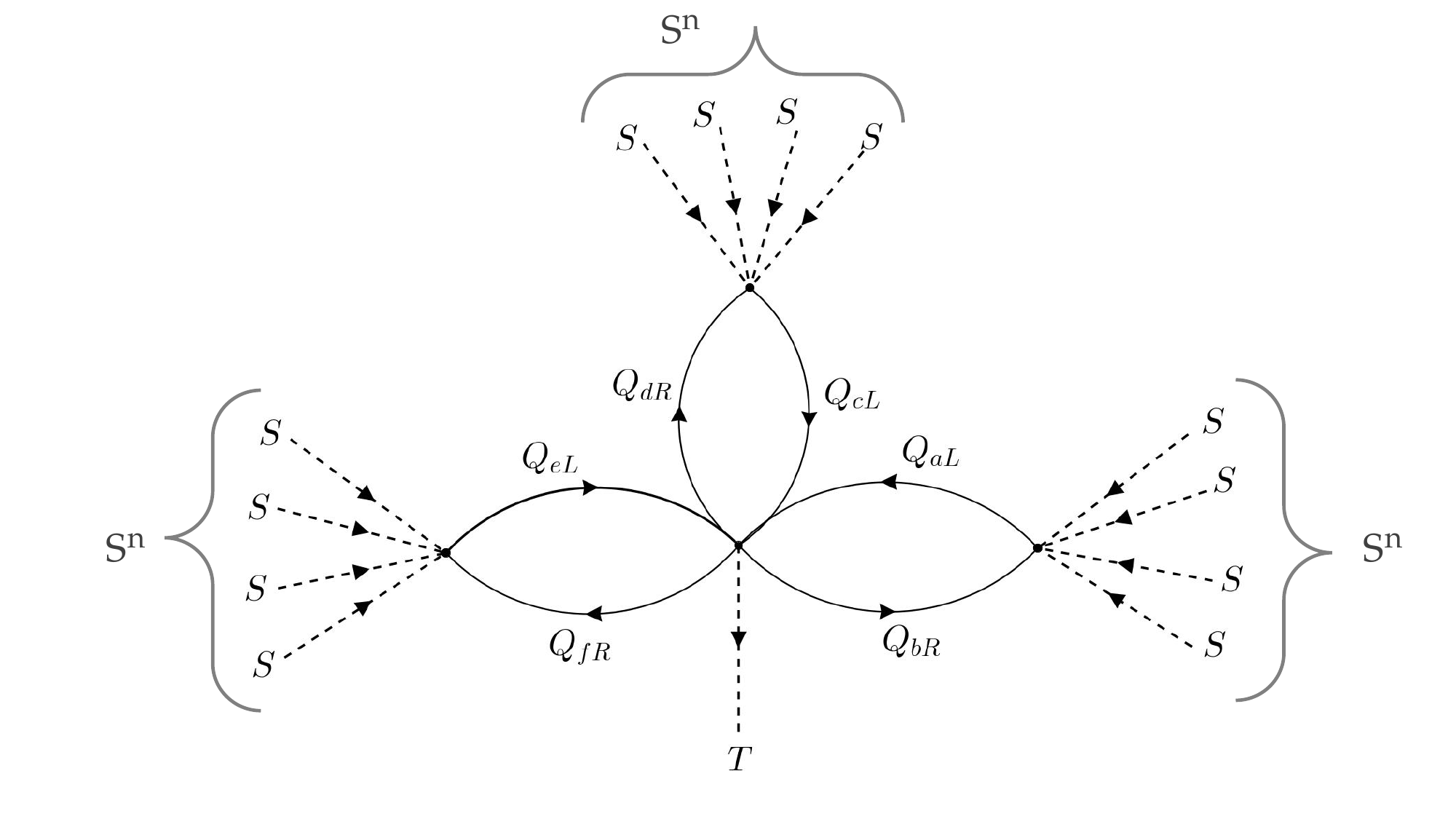}
	\caption{A three-loop diagram inducing a purely scalar operator $S^{3n}\,T^*$ that affects the axion quality. The relevant operators inducing this diagram are given in Eq. (\ref{eq:opmod1}) and Eq.~(\ref{eq:Yuk1}).}
	\label{fig:3loopmod1}
\end{figure}

There are four types of higher dimensional operators involving the VLQs that can destabilize the axion quality through loop diagrams.\footnote{We thank Vasja Susič for emphaszing to us the importance of loop diagrams for axion quality.} They are given by
\begin{eqnarray}
{\cal L}^{\rm Yuk}_{\rm gravity} &=&
\kappa_{ab} \frac{(\overline{Q}_{aL} Q_{bR})^3\, T}{M_{\rm Pl}^{6}} + \kappa_{44} \frac{\overline{Q}_{4L} Q_{4R}\, (S^*)^{3n}}{(3n)!\,M_{\rm Pl}^{3n-1}}\nonumber\\
&+& \kappa_{a4}\frac{\overline{Q}_{aL}Q_{4R}\,(S^*)^n}{n!\,M_{\rm Pl}^{n-1}}+ \kappa_{4a}\frac{\overline{Q}_{4L} Q_{aR}\,(S^*)^n}{n!\, M_{\rm Pl}^{n-1}} + h.c.
\label{eq:opmod1}
\end{eqnarray}
The first operator in Eq. (\ref{eq:opmod1}), when combined with the second term of Eq.~(\ref{eq:Yuk1}), would induce a purely scalar operator $S^{3n} \,T^*$ through a three-loop diagram shown in Fig.~\ref{fig:3loopmod1}.  
The induced Lagrangian can be estimated to be
\begin{equation}
{V}_{\rm gravity} \sim \kappa_{ab} \left(\frac{{\rm Tr}(Y^{(Q)})}{16 \pi^2}\right)^3 \frac{S^{3n}\, T^*}{(3n)!\, M_*^{3n-3}} \left(\frac{M_*}{M_{\rm Pl}} \right)^6 + h.c.
\label{eq:pot4}
\end{equation}
Each loop has a naive quadratic divergence, which is cut-off by the scale $M_*$.  
Comparing Eq.~(\ref{eq:pot4}) with Eq.~(\ref{eq:pot3}) we see that these loop-induced corrections can be more important, with some of the $M_{\rm Pl}$ factors replaced by $M_*$ in the denominator.  However, these contributions are safe as can be seen by setting $n=4$, $M_* = 1 \times 10^{12}$ GeV, Tr($Y^{(Q)}) = 0.1$, $\kappa_{ab}= 1$, in which case $\overline{\theta} \sim 1.2 \times 10^{-13}$.  For $n=5$ or larger, these corrections are even smaller. For example, for $n=5$, we find the induced $\overline{\theta}$ to be $\overline{\theta} \sim 1.2 \times 10^{-16}$ with the other parameters being the same as quoted above.

\begin{figure}[h!]
		\centering
		\includegraphics[width=0.9\textwidth]{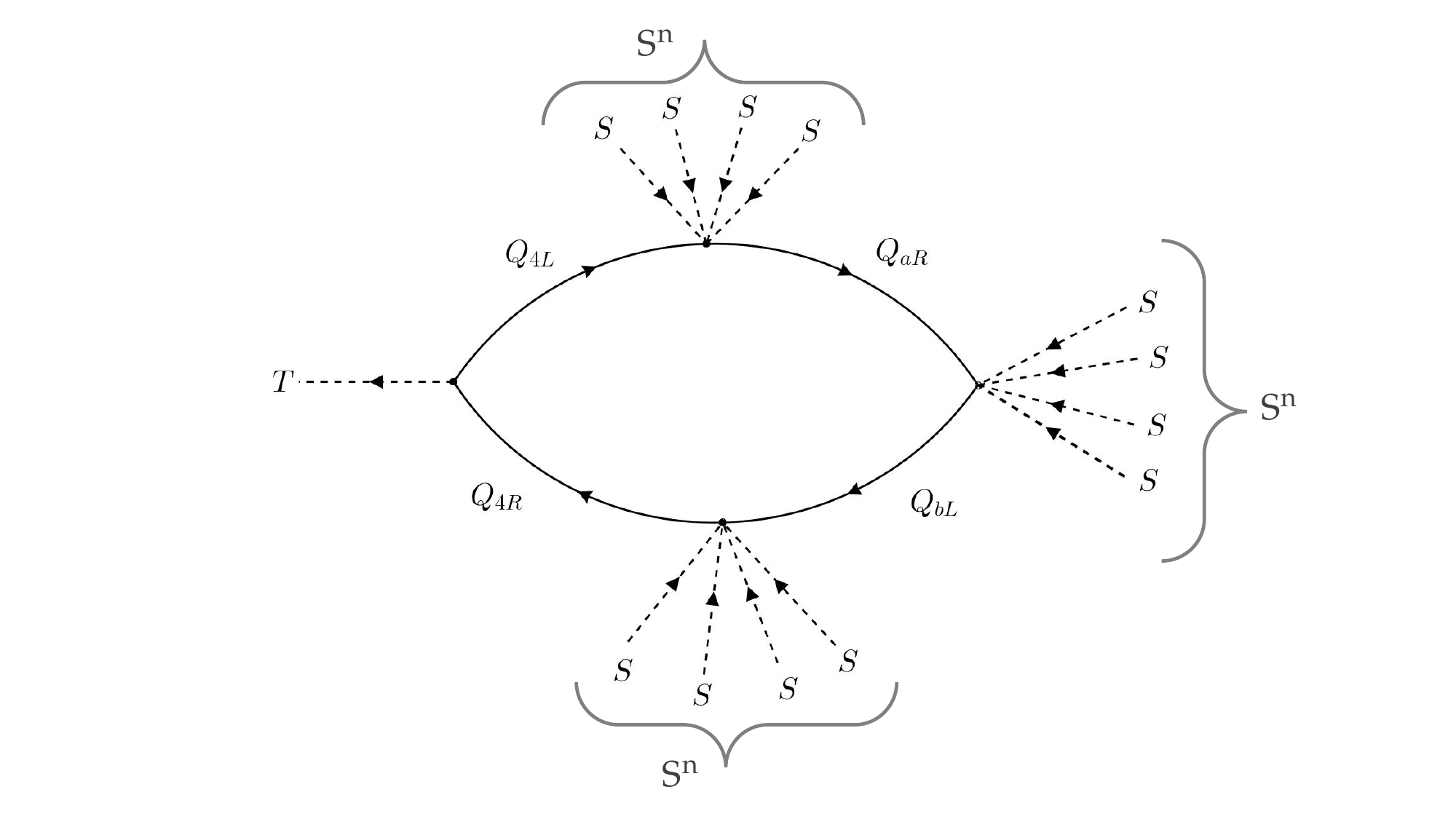}
	\caption{A one-loop diagram inducing a purely scalar operator that breaks the PQ symmetry.  This diagram uses the operators given in Eq.~(\ref{eq:opmod1}).}
	\label{fig:oneloopmod1}
\end{figure}

In Fig.~\ref{fig:oneloopmod1} we have shown a one-loop diagram that induces the operator $S^{3n}\,T^*$ by utilizing the last two terms of Eq.~(\ref{eq:opmod1}), along with the first term of Eq.~(\ref{eq:Yuk1}).  The induced potential arisng from this diagram can be estimated to be
\begin{equation}
V_{\rm gravity} \sim \frac{Y_{44}^{(Q)}}{16\pi^2}\left(\kappa_{a4}^* Y^{(Q)}_{ab} \kappa_{4b}^*  \right) \frac{S^{3n}\,T^*}{(3n)!\,M_{\rm Pl}^{3n-3}} \left(\frac{M_{\rm Pl}} {M_*}  \right)^{n-1}~.
\end{equation}
The associated shift in $\overline{\theta}$ is within acceptable limits for a range of parameters.  For example, with $n=4$, $f_T = f_S$, $f_a = 5 \times 10^{10}$ GeV, $M_* = 4 \times 10^{12}$ GeV, $Y^{Q)} = Y_{44}^{(Q)} =  0.05$, one would have $\overline{\theta} \sim 10^{-10}$.  Other acceptable choices of parameters are also possible. If $n=5$ is chosen, with the other parameters as quoted above, $\overline{\theta} \sim 10^{-28}$, well below the required value for a high quality axion.\footnote{If only three VLQs with axial charges $\pm(1,1,-2)$ are used, an effective operator $(\overline{Q}_{aL} Q_{bR})^2 T/M_{\rm Pl}^3$ would be induced by gravity.  In this case, a two-loop diagram analogous to Fig.~\ref{fig:3loopmod1} would induce a $\overline{\theta}$ that exceeds the allowed range. Thus we see that the minimum number of VLQs needed in this setup is 4.}

As for the the $\kappa_{44}$ term in Eq.~(\ref{eq:opmod1}), it would induce a sub-dominant shift in $\overline{\theta}$ owing to the higher power of $M_{\rm Pl}$ in the denominator of this term.  The effective scalar operator that arises from a one-loop diagram anaalogus to Fig.~\ref{fig:oneloopmod1} is given by
\begin{equation}
V_{\rm gravity} \sim \frac{Y^{(Q)}_{44}}{16 \pi^2}
\frac{\kappa_{44}\, S^{3n} T^*}{(3n)!\, M_{\rm Pl}^{3n-3}}\left(\frac{M_*}{M_{\rm Pl}} \right)^2 + h.c.
\label{eq:pot4}
\end{equation}
Comparing Eq.~(\ref{eq:pot4}) with Eq.~(\ref{eq:pot3}) one finds that this contribution is suppressed by a loop factor as well as an additional factor $(M_*/M_{\rm Pl})^2 \ll 1$. Thus, once the condition of Eq.~(\ref{eq:pot3}) is satisfied, so is the condition of Eq.~(\ref{eq:pot4}).

In the gravity-induced operators we have assumed the presence of a $n!$ factor in the denominator for $n$ identical particles, in analogy with quantum field theory calculations. If such factors are not present, while the $n=4$ case would lose its high-quality axion, any larger value of $n \geq 5$ will still preserve its high quality.

\subsection{Axion coupling to fermions and photon in Model I}
By construction, model I yields a KSVZ-type axion. Here and in what follows  we shall follow the standard convention to describe axion interaction with matter and radiation, which takes the form~\cite{DiLuzio:2020wdo}
\begin{equation}
{\cal L}^{\rm int}_a \supset \frac{\alpha_s}{8 \pi} \frac{a}{f_a}G_{\mu \nu}^a \tilde{G}^{a,\mu \nu} + \frac{\alpha}{8 \pi} \frac{C_{a\gamma}}{f_a} F_{\mu \nu} \tilde{F}^{\mu \nu} + C_{af} \frac{\partial_\mu a} {2 f_a}(\overline{f} \gamma^\mu \gamma_5 f)~.
\label{eq:acoup}
\end{equation}
where $f=e,p,n$.  The coefficients of various terms appearing in Eq. (\ref{eq:acoup}) are given by~\cite{DiLuzio:2020wdo}
\begin{eqnarray}
C_{a\gamma} &=& \frac{E}{N} - 1.92 \nonumber \\
C_{ap} &=& -0.47+0.88 \,c_u^0 -0.39 \,c_d^0-C_{a,{\rm sea}} \nonumber \\
C_{an} &=& -0.02 +0.88 \,c_d^0-0.39 \,c_u^0-C_{a,{\rm sea}} \nonumber \\
C_{a,{\rm sea}}  &=& 0.038\, c_s^0+0.012 \,c_c^0+0.009 \,c_b^0+0.0035 \,c_t^0 \nonumber \\
C_{ae} &=& c_e^0 + \frac{3 \alpha^2}{4\pi^2} \left[\frac{E}{N}\,{\rm log}\left(\frac{f_a}{m_e}\right) -1.92\, {\rm log}\left(\frac{\rm GeV}{m_e}\right)\right]~.
\label{eq:acoup2}
\end{eqnarray}
Here the factor $E$ and $N$ respectively are the electromagnetic and color anomaly coefficients arising from the triangle diagrams that convert the axion to two photons and two gluons. The factor $-1.92$ in $C_{a\gamma}$ arises from non-perturbative effects which includes $a-\pi^0$ mixing. 
The coefficients $c_i^0$ in Eq, (\ref{eq:acoup2}) are model-dependent direct couplings of the axion.  The coefficients $-0.47$ in $C_{ap}$ and $-0.02$ in $C_{an}$ are universal to all axion models, arising from the first coupling of Eq. (\ref{eq:acoup}). The second term in $C_{ae}$ arises from the two-photon coupling of $a$, where the photons are converted to electrons via a loop diagram.

The last two terms of Eq. (\ref{eq:acoup}) are also written sometimes as
\begin{equation}
{\cal L}_a^{\rm int} \supset \frac{1}{4}g_{a\gamma} F_{\mu \nu}\tilde{F}^{\mu\nu} + g_{af} \frac{\partial_\mu a}{2m_f}(\overline{f} \gamma^\mu \gamma_5f)
\label{eq:acoup3}
\end{equation}
leading to the correspondence
\begin{equation}
g_{a \gamma}= \frac{\alpha}{2 \pi}\frac{C_{a\gamma}}{f_a},~~~g_{af} = C_{af} \frac{m_f}{f_a}~.
\label{eq:acoup4}
\end{equation}

In model I, the axion does not couple to the SM fermions at tree level.  As such, all the $c_i^0$ coefficients in Eq. (\ref{eq:acoup2}) are zero.  
Axion couplings to the nucleon arise from the $aG \tilde{G}$ term which are universal to all models. As for its coupling to photons, due to the non-zero hypercharge $y$ of the VLQs of the model, there is a finite contribution. We find $E/N = 8/3$ if $y=2/3$ is chosen, and $E/N = 2/3$ if $y=-1/3$ is used.  The value of $E/N=8/3$ is identical to the value obtained in the DFSZ axion model~\cite{DiLuzio:2020wdo}. If $y=0$ is chosen,  there is a small three-loop contribution to $g_{a\gamma}$, with the first loop connecting $a$ to gluons via the $Q$ fermions and the gluons in turn connecting to SM quarks which then give rise to two photons. This coupling is however suppressed by an additional factor of $(\alpha \alpha_s)/(16\pi^2)\sim 2 \times 10^{-5}$, compared to the other scenarios with $y \neq 0$. These properties of the model, while difficult to disentangle from the KSVZ model, can potentially be used as tests of the model.

In the case that the hyercharges $y$ of the VLQs is chosen so that $y=2/3$ or $y=-1/3$, operators of the type $\overline{Q}_{aL} u_R S^{n/2}/M_{\rm Pl}^{n/2-1}$ or 
$\overline{Q}_{aL} d_R S^{n/2}/M_{\rm Pl}^{n/2-1}$ (for even $n$), can arise from gravity that can lead to mixing of the VLQs with the SM quarks $(u,\,d)$, and thereby their decays. This would evade cosmological constraints for a stable vector-like quark. Such operators do not affect the axion quality, as the SM sector can be grouped together with the scalar sector $S$.

\subsection{Domain wall number in Model I} 

Let us now comment on the question of domain wall number and potential cosmological problems associated with them in this model~\cite{Sikivie:1982qv}. 
Since the global PQ symmetry is not uniquely defined in the theory -- any linear combination of a PQ symmetry and the gauged $U(1)_a$ symmetry also is a valid PQ symmetry -- calculation of the domain wall number $N_{\rm DM}$ requires more care.  We follow the procedure developed in Ref.~\cite{Ernst:2018bib} to dertimen $N_{\rm DW}$. Once the axion field $a$ is identified as in Eq.~(\ref{eq:axionmod1}) in terms of the pseudoscalar fields contained in the original fields, one can write in general
\begin{equation}
a= \sum_i c_i \,\eta_i~.
\label{eq:ci}
\end{equation}
The coefficients $c_i$ for model I can be read off from Eq.~(\ref{eq:axionmod1}).  
We have  $c_S = 3 f_T n/\sqrt{9 f_T^2 n^2 + f_S^2}$, $c_T = -f_S/\sqrt{9 f_T^2 n^2 + f_S^2}$, with the decay constant $f_a$ given as in Eq.~(\ref{eq:famod1}).

The domain wall number can be computed from the formula~\cite{Ernst:2018bib}
\begin{equation}
N_{\rm DW} = {\rm minimum ~ integer} \left\{\frac{1}{f_a} \sum_i n_i\, c_i\, f_i~,~~~n_i \in {\cal Z} \right\}~.
\label{DW}
\end{equation}
Using the values of $c_S$, $c_T$, and $f_a$ for the model,  we obtain 
\begin{equation}
N_{\rm DW} = {\rm minimum ~ integer} (3 n_1 n - n_2)~.
\label{eq:rhs}
\end{equation}
One should choose the integers $n_1$ and $n_2$ so as to find the minimum value for the right-hand side of Eq. (\ref{eq:rhs}). Taking $n_1 = 1,\, n_2 = 3 n-1$, we find $N_{\rm DW} = 1.$ It is interesting that even though the model has 4 VLQs, the domain wall number turns out to be 1.  This is a consequence of the interplay between the $U(1)_{\rm PQ}$ and the gauged $U(1)_a$.  Domain wall number of one does not cause any problem with early universe cosmology, since it is unstable and dissipates well above the BBN temperature.  It does contribute to the relic density of the axion, which we shall take into account in constraining the axion mass and $f_a$.

\subsection{Axion as dark matter in Model I}

In a post-inflationary PQ breaking scenario, which is what we adopt, the axion can be the entire dark matter (DM) component of the universe.  The needed value of the axion mass and decay constant has been investigated numerically in Ref.~\cite{Kawasaki:2014sqa}, which finds a range 
\begin{equation}
f_a = (4.6-7.2) \times 10^{10}~{\rm GeV}
\label{eq:fa1}
\end{equation}
for the net contribution to the axion relic density to be consistent with observations corresponding to the case of $N_{\rm DW} = 1$.  This range corresponds to the axion being the entire dark matter component of the Universe, with 
$\Omega_a h^2 \simeq0.12$.  
These contributions include three sources of axion production: via vacuum misalignment during the QCD phase transition, axions from global string decays, and axions from the decay of the domain wall.  The axion mass corresponding to the range of $f_a$ quoted in Eq.~(\ref{eq:fa1}) is (see Eq. (\ref{eq:axionmass})) 
\begin{equation}
m_a = (80-130)~\mu eV~.
\end{equation}
A more recent simulation~\cite{Benabou:2024msj} finds a slightly wider range of the axion mass, $m_a = (40-300)~\mu eV$.  A separate simulation of Ref.~\cite{Gorghetto:2018myk} quotes larger uncertainties and finds $m_a \gtrsim 500~\mu eV$. 
Since the quality of the axion is quite good within the model, both ranges of the axion mass are fully consistent, and the model can easily accommodate the entire DM in the form of axion. 

\subsection{A UV complete model for the mass scale $M_*$}

\begin{figure}[t!]
		\centering
		\includegraphics[width=1.0\textwidth]{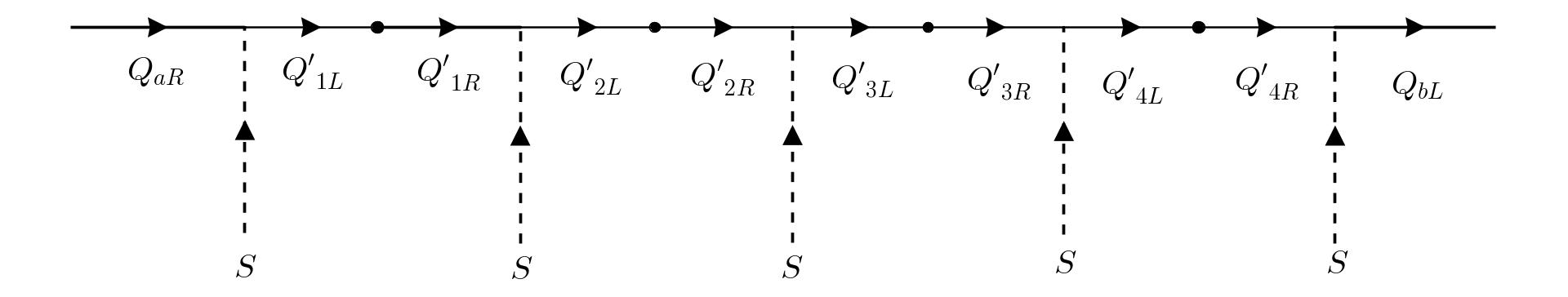}
	\caption{A UV-complete diagram to induce the VLQ masses in Model I.  Here we have adopted $n=5$. The primes quarks have vectorial charges under $U(1)_a$, allowing for their bare masses, which are indicated as dots in the figure.}
    \label{fig:UV}
\end{figure}

The Yukawa couplings of Eq.~(\ref{eq:Yuk1}) make use of higher dimensional operators suppressed by powers of a scale $M_*$ to induce the masses of the VLQs with $U(1)_a$ charges $\pm 1$.  If $M_*$ is identified as $M_{\rm Pl}$, with the integer $n$ obeying  $n\geq 4$ to realize a high-quality axion, these VLQ masses would be well below a TeV which would conflict direct experimental limits.  This implies that some form of new physics should lie below $M_{\rm Pl}$ that induces the VLQ masses. Here we show an example, which would provide a UV completion of the model. 

One way of UV-completing the model is presented in Fig.~\ref{fig:UV}.  This corresponds to the case of $n=5$, so that the effective operator involving the $Q_{a}$ fields is $\overline{Q}_{aL} Q_{bR} S^5/M_*^4$.  Here $M_*$ may be identified as the (common) mass of the VLQs labeled $Q'$ which have vectorial $U(1)_a$ charges.  These new fermions, with vectorial $U(1)_a$ assignment, will not modify the anomaly cancellation conditions for the $U(1)_a$ sector. 
 Specifically, for $n=5$ we take the charges of these $Q'$ quarks to be 
\begin{equation}
Q'_{1}: -\frac{3}{5},~~Q'_{2}: -\frac{1}{5},~~
Q'_{3}: \frac{1}{5},~~Q'_{4}: -\frac{3}{5},~~{\rm and}~~Q_{aL}: 1,~~Q_{bR}:-1,~~S: \frac{2}{5}~.
\label{eq:chargeUV}
\end{equation}
With these charges, the following renormalizable Yukawa couplings can be written down:
\begin{eqnarray}
{\cal L}_{\rm Yuk}^{\rm UV} &=& \overline{Q}_{bL} Q'_{4R} S + \overline{Q'}_{4L} Q'_{3R} S + \overline{Q'}_{3L} Q'_{2R} S + \overline{Q'}_{2L} Q'_{1R} S \nonumber \\
&+&  \overline{Q'}_{1L} Q_{aR} S 
+ \sum_{\alpha=1}^4 M_\alpha \overline{Q'}_{\alpha L} Q'_{\alpha R} + h.c.
\end{eqnarray}
We see that the effective operator $\overline{Q}_{bL} Q_{aR}S^4$ will be induced, once the $Q'$ quarks are integrated out.

Once the UV-completion of the model is achieved, it is important to check if the new fields introduced can have an effect on the axion quality.  With the charges of the $Q'$ fields as given in Eq.~(\ref{eq:chargeUV}), one can consider the fields  $(Q_{aL}, Q_{aR}, S, Q')$ to belong to one sector which preserves the global $U(1)_{\rm PQ}$ symmetry.  Operators of the type $\overline{Q'}_{1L} Q_{4R} (S^*)^6/M_{\rm Pl}^5$ will be induced by gravity, but such couplings do not upset the axion quality. These couplings have to be combined with other such couplings, which would make the resulting shift in $\overline{\theta}$ well within the allowed range.

As for the $N_3$ mass term in Eq. (\ref{eq:Yuk1}), the corresponding $M_*$ may actually be $M_{\rm Pl}$, in which case $M_{N_3} \sim f_a^2/M_{\rm Pl} \sim {\rm TeV}$.  Alternatively, this mass term can arise by integrating out a singlet fermion $N_0$ with zero $U(1)_a$ charge.

\section{  Model II: A family of models generalizing  Model I}

\label{sec:Model2}

There is a simple generalization of model I to a family of models.  Here the number of vector-like quarks is
chosen to be $m + 1$ with $m$ of them carrying $U(1)_a$ 
charge of $+1$ (for left-handed sector) and $-1$ (for right-handed sector) while the $(m+1)^{\rm th}$ quark has charge $\mp m$. This assignment, shown in Table~\ref{tab:tab2}, is anomaly-free, including contributions from three singlet fermions $(N_{R})_{1,2,3}$ with their charges as indicated. This provides a family of models, which is anomaly-free for arbitrary integer values of $m$.  As in Model I, two SM singlet scalars, $T$ and $S$ are introduced with $U(1)_a$ charges of $2m$ and $2/n$ respectively. Here $n$ is a positive integer with its allowed values for axion quality depending on the integer $m$. The choice $m=3$ corresponds to Model I.

\begin{table}[h] 
   \label{tab:example}
   \small
   \centering
   \begin{tabular}{|c|c|c|}
   \hline
   \textbf{New fermion} & \textbf{\boldmath{$SU(3)_c \times SU(2)_L \times U(1)_Y$} } & \textbf{\boldmath{$U(1)_a$} charge}  \\ 
   \hline
   $(Q_L)_{1,2,3,...,m,m+1}$& $(3,1,y)$ & ~$(1,1,1,..1,-m)$ \\
   \hline
   $(Q_R)_{1,2,3,...,m,m+1}$ & $(3,1,y)$ & $-(1,1,1,...1,-m)$ \\
   \hline
     $(N_R)_{1,2,3}$& $(1,1,0)$ & ~$(-1+m,1+m,-2m)$ \\
   \hline
   \textbf{New scalar} & \textbf{\boldmath{$SU(3)_c \times SU(2)_L \times U(1)_Y$} } & \textbf{\boldmath{$U(1)_a$} charge}\\
\hline
$T$& $(1,1,0)$ & $2m$ \\
   \hline
   $S$ & $(1,1,0)$ & $2/n$ \\
   \hline\hline 
   \end{tabular}
   \caption{New fermions and scalars along with their transformation under the standard model and their respective  $U(1)_a$ charges in Model II. Here $m \geq 3$ and $n$ are positive integers. If $m=3$ is chosen, this model reduces to Model I with charges given in Table \ref{tab:tab1}. The allowed value of $n$ varies, and depends on the value of $m$ chosen.}
   \label{tab:tab2}
\end{table}

The explicit calculations for the case of $m+1$ quark fields follow along the lines shown for model I ($m=3$). The Yukawa Lagrangian in this case is given by
\begin{equation}
   - {\cal L}_{\rm Yuk} = Y_{m+1} \,\overline{Q}_{m+1,L} Q_{m+1,R} T^* + \sum_{i=1}^mY_{i} \,\overline{Q}_{iL} Q_{iR}\, \frac{S^n}{(M_*)^{(n-1)}} + h.c.
\end{equation}
The Goldstone field eaten up by the $U(1)_a$ gauge boson and the orthogonal and axion field are given by
\begin{eqnarray}
G = \frac{\left(\frac{2}{n}f_S \eta_S + 2m f_T \eta_T  \right)}{\sqrt{(2m f_T)^2+ (\frac{2}{n}f_S)^2}},~~~a = \frac{\left(2 m f_T \eta_S -\frac{2}{n}f_S \eta_T  \right)}{\sqrt{(2m f_T)^2+ (\frac{2}{n}f_S)^2}}~.
\end{eqnarray}

The Yukawa Lagrangian involving the axion and the VLQs is now
\begin{equation}
    -{\cal L}_{\rm Yuk}(a) = i a \left[ \overline{Q}_{m+1} \gamma_5 Q_{m+1} \frac{f_S}{f_T} \frac{M_{Q_{m+1}}}{ \sqrt{f_T^2 m^2n^2 + f_S^2}} + \sum_{i=1}^m\overline{Q}_{i} \gamma_5 Q_{i}\frac{m n^2 f_T}{f_S} \frac{M_{Q_{i}}}{\sqrt{f_T^2 m^2 n^2 + f_S^2}}
    \right]~.
\end{equation}
Summing over the $(m+1)$ quarks in computing $a\,G \tilde{G}$ coefficient, one obtains
\begin{equation}
\frac{1}{f_a} = \frac{f_S}{f_T}\frac{1}{\sqrt{f_T^2 m^2 n^2 + f_S^2}} + (m \times m n^2) \frac{f_T}{f_S} \frac{1}{\sqrt{f_T^2 m^2 n^2 + f_S^2}}~.
\end{equation}
This simplifies to
\begin{equation}
f_a = \frac{f_T f_S}{\sqrt{f_T^2 m^2 n^2 + f_S^2}}~.
\label{eq:fa2}
\end{equation}

In this model we have $c_S = m n f_T/\sqrt{f_T^2 m^2 n^2 + f_S^2}$, $c_T =-f_S/\sqrt{f_T^2 m^2 n^2 + f_S^2}$ for the $c_i$ factors defined in Eq. (\ref{eq:ci}). Thus, using the value of $f_a$ given in Eq. (\ref{eq:fa2}) we get
\begin{equation}
    N_{\rm DW} = {\rm minimum~ integer} ( m n n_1 - n_2)~.
\end{equation}
Choosing $n_1 = 1, \, n_2 = m n-1$ gives the minimum (positive) integer which is $N_{\rm DW} = 1$.
Thus, this family of models are free of cosmological difficulties associated with domain walls.


\vspace*{0.1in}
\subsection{Axion quality in Model II }

A gravity-induced operator involving the scalar fields $T$ and $S$ 
\begin{equation}
V_{\rm gravity} = \frac{\kappa e^{i \delta}}{(mn)!} \frac{S^{mn} T^*}{M_{\rm Pl}^{mn-3}} + h.c.
\end{equation}
can lead to a shift in $\overline{\theta}$.
This term would lead to an axion potential given by
\begin{equation}
V_{\rm gravity}^{(a)} = \frac{f_S^{mn}f_T \kappa}{2^{(mn-1)/2} (mn)! M_{\rm Pl}^{mn-3}} \cos\left(\frac{a}{f_a}+\delta\right)~.
\end{equation}
Minimizing this potential, along with that of Eq. (\ref{eq:axionpot2}), will induce a $\overline{\theta}$ given by
\begin{equation}
\overline{\theta} \simeq \frac{\kappa \sin\delta}{2^{(mn-1)/2}(mn)!} \frac{(m_u+m_d)^2}{m_u m_d} \frac{1}{f_\pi^2 m_\pi^2} \frac{f_S^{mn}f_T}{M_{\rm Pl}^{mn-3}}~.
\end{equation}
This equation reduces to Eq.~(\ref{eq:shiftmod1}) for the case of $m=3$.  For $m > 3$. $\overline{\theta}$ is more suppressed than for the case of $m=3$. For example, for $m=4$ and with $f_S = f_T = f_a\sqrt{1+m^2n^2}$ we find
\begin{equation}
  |\overline{\theta}| \simeq
    \begin{cases}
      1.2 \times 10^{-57} ~~ (n=4,\,f_a = 5 \times 10^{10}~{\rm GeV})\\
      2.7 \times 10^{-25} ~~(n=3, \, f_a = 5 \times 10^{10}~{\rm GeV})~.
    \end{cases}       
\end{equation}
In this case, axion quality allows for a lower value of the integer, $n=3$, compared to $n=4$ that was used in Model I.  

As for loop-induced corrections to $\overline{\theta}$, there are four sets of terms involving VLQs, analogous to Eq.~(\ref{eq:opmod1}). They are given by
\begin{eqnarray}
&~&{\cal L}^{\rm Yuk}_{\rm gravity} =
\kappa_{ab} \frac{(\overline{Q}_{aL} Q_{bR})^m\, T}{M_{\rm Pl}^{6}} + \kappa_{m+1,m+1} \frac{\overline{Q}_{m+1,L} Q_{m+1,R}\, (S^*)^{mn}}{(mn)!\,M_{\rm Pl}^{mn-1}}\nonumber\\
&~& + \kappa_{a,m+1}\frac{\overline{Q}_{aL}Q_{m+1,R}\,(S^*)^{\frac{n(m-1)}{2}}}
{(\frac{n(m-1)}{2})!\,M_{\rm Pl}^{\frac{n(m-1)}{2}-1}}+ 
\kappa_{m+1,a}\frac{\overline{Q}_{m+1,L}Q_{aR}\,(S^*)^{\frac{n(m-1)}{2}}}
{(\frac{n(m-1)}{2})!\,M_{\rm Pl}^{\frac{n(m-1)}{2}-1}} + h.c.
\label{eq:opmod2}
\end{eqnarray}
Here the last two terms are allowed only when $n(m-1)/2$ is an integer. The case of $m=3$ reduces to Eq. (\ref{eq:opmod1}).  These operators will lead through loops analogous to Fig.~\ref{fig:3loopmod1} and Fig.~\ref{fig:oneloopmod1} to purely scalar operators of the form $S^{mn} T^*$.  However, these effects are much more suppressed for $m > 3$ compared to the case of $m=3$.  For example, the equivalent of Fig.~\ref{fig:3loopmod1} would have $m$ loops.  We conclude that for $m > 3$ the axion quality constraint is easily satisfied.


\subsection{A special case of \boldmath{$m+1 \geq 10$} and \boldmath{$n=1$} in Model II}

The case of $m+1 \geq 10$ in model II allows for the choice $n=1$ (see charges listed in Table~\ref{tab:tab2}).  With this choice, the Yukawa couplings of all vector-like quarks are allowed at the renormalizable level.  They are given as
\begin{equation}
   - {\cal L}_{\rm Yuk} = Y_{m+1} \,\overline{Q}_{m+1,L} Q_{m+1,R} T^* + \sum_{i=1}^mY_{i} \,\overline{Q}_{iL} Q_{iR}\, S + h.c.
\end{equation}
This subclass of Model II is very interesting since it is already UV complete.  The lowest order gravity-induced $U(1)_{\rm PQ}$ symmetry breaking term among scalars is $S^m T^*/M_{\rm Pl}^{m-3}$.  For $m\geq 9$, these operators have at least six powers of $M_{\rm Pl}$ suppression, which are safe from destabilizing the axion solution.

This special case of models is in some sense analogous to the Barr-Seckel models~\cite{Barr:1992qq}.  Both require at least ten vector-like fermions for axion quality. Both are renormalizable models.  These are, however, different models with differing predictions for certain observables.  In our subclass of model II, the domain wall number is always one, while it can be different from one in the Barr-Seckel models. In Appendix~\ref{sec:Appendix-A}, we have briefly reviewed the Barr-Seckel framework for comparison, where we also discuss the possibility of addressing the cosmological domain wall problem in that setup by relying on the gravity-induced operators.

\section{Model III: A high quality \boldmath{$SO(10)$} axion model}

\label{sec:Model3}

The similarity of the seesaw scale for neutrino mass generation and the invisible axion scale has over the years, inspired many works on the non-supersymmetric $SO(10)$ as well as other models for the invisible axion ~\cite{Mohapatra:1982tc,Babu:2015bna,Bertolini:2014aia,Babu:2018qca,Ernst:2018bib,DiLuzio:2018gqe,Ballesteros:2019tvf,Boucenna:2018wjc, Dev:2018pjn,Bertolini:2020hjc,Lazarides:2022ezc}. These works, however, do not address the problem of Planck scale effects and hence they leave the axion quality problem untouched. In this section, we present a realistic grand unified model based on the $SO(10)$ gauge group which solves the quality problem in a similar way to models I and II presented here, by making use of an axial $U(1)_a$ gauge symmetry. A brief account of the results in this section has been presented in Ref.~\cite{Babu:2024udi}. Here we provide further details and also extend our analysis to include loop-induced effects that may affect the axion quality.

The fermions and scalars of the $SO(10) \times U(1)_a$ gauge model are listed in Table~\ref{tab:tab3} where their representations and $U(1)_a$ charges are shown. The fermion sector of the model includes three copies of the SM fermions grouped under ${\bf 16}_a$ ($a=1-3$), as well as a single {\bf 10} and four $SO(10)$ singlet fermions ($\chi,\,N_a)$.  This choice allows for realizing $U(1)_a$ as a gauge symmetry in a simple way.  All gauge anomalies cancel, as can be seen from these conditions: 
\begin{eqnarray}
A[SO(10)^2 \times U(1)_a] &=& 3 \times 2 \times 1 + 1 \times 1 \times (-6) = 0 \nonumber \\
A[{(\rm gravity})^2 \times U(1)_a] &=& 3 \times 16 \times 1 + 1 \times 10 \times (-6) + 1 \times 1 \times 12 + 2 \times 1 \times (-4)\nonumber \\
&+& 1 \times 1 \times 8 = 0 \nonumber\\
A[(U(1)_a)^3] &=& 3 \times 16 \times (1)^3 + 1 \times 10 \times (-6)^3 + 1 \times 1 \times (12)^3 + 2 \times 1 \times (-4)^3 \nonumber \\
&+& 1 \times 1 \times (8)^3 = 0~.
\end{eqnarray}

The $U(1)_a$ charge assignment shown in Table \ref{tab:tab3} is arguably the simplest in an $SO(10) \times U(1)_a$ gauge theory.  Suppose we assign a family-universal charge of $+1$ to the ${\bf 16}_i$ in an $SO(10) \times U(1)_a$ gauge theory. $SO(10)^2 \times U(1)_a$ anomaly cancellation would require a new $SO(10)$ non-singlet fermion, with the simplest choice being a single {\bf 10}, with its charge being $-6$, as shown in Table \ref{tab:tab3}. (Note that the index of ${\bf 16}$ is 2, while that of {\bf 10} is 1.)   The $({\rm gravity})^2 \times U(1)_a$ anomaly receives contributions of $3 \times 16 \times 1 + 10 \times (-6) = -12$ from the $SO(10)$ non-singlet fermions.  Similarly, the $[U(1)_a]^3$ anomaly receives a contribution of $3 \times 16 \times 1^3 + 10 \times (-6)^3 = -2112$ from these fermions.  Additional fermions are therefore needed for anomaly cancellation with $U(1)_a$ charges $Q_i$ satisfying 
\begin{equation}
\sum_iQ_i = +12,~~\sum_i(Q_i)^3 = + 2112~.
\end{equation}
With a single $SO(10)$ singlet fermion added, there is no solution to these equations.  If two fields added, the solutions lead to irrational charges, which we exclude.  We have not found a solution to these equations with three fermion fields based on our numerical searches.  Thus, a minimum of four fermion fields are needed. With the fields $N_a$ and $\chi$ charges as indicated in Table \ref{tab:tab3}, there is a solution, which is the simplest possibility.

Also listed in Table~\ref{tab:tab3} are the charges of various fields under a global $U(1)$, which is accidental within the model.  These global $U(1)$ charges are not uniquely determined, since any linear combination of a global $U(1)$ and the gauged $U(1)_a$ is also an equally good global symmetry.

\begin{table}[htbp]
\begin{center}
\begin{tabular}{|c||c||c||c|}\hline 
\textbf{Fermion} & \textbf{\boldmath{$SO(10)$}} & \textbf{\textbf{gauge} \boldmath{$U(1)_a$}} & \textbf{global} \boldmath{ $U(1)$}\\
& \textbf{irrep} & \textbf{charge} &\textbf{charge}\\\hline
$\psi_a$ & {\bf 16}$_a$ & +1 & +1\\
$F$ & {\bf 10} & $-6$ & ~~0\\
$\chi$ & {\bf 1} & $+12$ & ~~0\\
$N_{1,2,3}$ & {\bf 1} &$(-4, -4, +8)$& $$(0,\,0,+2)$$  \\\hline\hline
\textbf{Scalar} & \textbf{\boldmath{$SO(10)$} rep} & \textbf{\boldmath{$U(1)_a$} charge} &  \textbf{global} \boldmath{$U(1)$} \\\hline
$H$ & {\bf 10} & $-2$ & $-2$ \\
$H'$ & {\bf 10} & ~~$0$ & ~~0\\
$\overline{\Delta}$ & $\overline{{\bf 126}}$& $-2$ & ~$-2$\\
$T$ & {\bf 1} & +1& ~+1\\
$S$ & {\bf 1} & +12 & ~~~0 \\
$A$ & {\bf 45/210} & ~0 & ~~~0\\\hline\hline
\end{tabular}
\end{center}
\caption{Fermion and scalar multiplets of the $SO(10) \times U(1)_a$ model that solves the axion quality problem. 
There are three copies of ${\bf 16}_a$ fermions, corresponding to three generations. All fermion fields are taken to be left-handed. The last column lists an accidental global $U(1)$ symmetry present in the model, with a QCD anomaly which will be identified as $U(1)_{\rm PQ}$. The multiplet $A$ could either be a {\bf 45} or a {\bf 210} of $SO(10)$. The Higgs 10-plet $H$ is complex, while $H'$ is real.
}
\label{tab:tab3}
\end{table}

The Yukawa Lagrangian of the model, consistent with the gauge symmetries, is given by
\begin{eqnarray}\nonumber
{\cal L}_{\rm Yuk} &=&\psi^T\left(Y_{10}\,H+Y_{126}\overline{\Delta}\,\right)\psi\
+ FFS + \chi\chi (S^*)^2/M_{\rm Pl}\\\nonumber &+&  FN_3 H+
 \chi N_{1,2}(T^{4}S^*/M^4_{\rm Pl}+T^{*8}/M^7_{\rm Pl})\nonumber\\&+&\sum_{a,b=1,2}N_aN_b(T^{*4}S/M^4_{\rm Pl}+ T^8/M^7_{\rm Pl}) + N_{1,2}N_3T^{*4}/M^3_{\rm Pl} +h.c.
\label{eq:Yuk}
\end{eqnarray}
In Eq. (\ref{eq:Yuk}) $Y_{10}$ and $Y_{126}$ are symmetric Yukawa coupling matrices that lead to a realistic and predictive fermion spectrum.  In particular, the model accommodates large neutrino mixing angles simultaneously with small quark mixing angles. The predictions of the Yukawa sector for neutrino oscillations and charged fermion masses have been analyzed extensively in a variety of papers~\cite{Babu:1992ia,Bajc:2001fe,Fukuyama:2002ch,Bajc:2002iw,Goh:2003sy,Goh:2003hf,Babu:2005ia,Bertolini:2004eq,Bertolini:2005qb,Bertolini:2006pe,Bajc:2008dc,Joshipura:2011nn,Dueck:2013gca,Altarelli:2013aqa, Fukuyama:2015kra,Babu:2018tfi,Ohlsson:2019sja,Babu:2020tnf}.

The $FFS$ term in Eq. (\ref{eq:Yuk}) induces a mass for the 10-plet fermion, which is needed for consistency.  Eq. (\ref{eq:Yuk}) also contains certain non-renormalizable operators that are induced by gravity with appropriate powers of $M_{\rm Pl}$ suppression, with the
fourth term inducing a TeV-scale mass for the singlet fermion $\chi$. The coupling $F N_3 H$ plays an important role in the decay of the color-triplet fermion $F({\bf 3})$ from the {\bf 10},  $F({\bf 3}) \rightarrow \overline{N}_3 + \overline{t} + \overline{b}$, through the exchange of the color-triplet scalar in $H$. The lifetime for this decay can be estimated to be of order $10^{-11}$ sec., with the mass of $F({\bf 3})$ being of order $10^{11}$ GeV and the mass of the color-triplet scalar being at the GUT scale, which is compatible with big bang nucleosynthesis (BBN) constraints. The remaining terms in Eq. (\ref{eq:Yuk}) lead to sub-eV masses for the singlet fermions $N_a$.

The most general Higgs potential for a similar model has been studied in Ref.~\cite{Babu:2015bna}. Here we write down the relevant terms symbolically, focusing on nontrivial invariants (for the case of $A={\bf 45}$), including the $SO(10)$ singlet fields ($T,\,S)$: 
\begin{eqnarray}\nonumber
\label{pot}
V &\supset& HH'T^2+
 H \Delta AA+ \Delta \Delta HH+\Delta \bar{\Delta} \Delta H  \nonumber \\&+& \Delta \bar{\Delta} HH^* 
 +\Delta \bar{\Delta}\Delta \bar{\Delta}+T^{12}S^*/M^9_{\rm Pl}+ h.c.
 \label{eq:pot1model3}
\end{eqnarray}
This potential leads to the mixing of Higgs doublets from {\bf 10} and {\bf 126}, which is needed for realistic fermion mass generation.  The first term in Eq. (\ref{eq:pot1model3}),  $H H' T^2$, is needed to avoid a weak-scale axion. The Higgs potential has an accidental global $U(1)_{\rm PQ}$ symmetry with the charges listed in the fourth column of Table~\ref{tab:tab3}. Note that this global $U(1)$ has a QCD anomaly, with a coefficient of $+6$. This global symmetry, however, is not respected by quantum gravity, and is broken explicitly by  the Planck-suppressed operators. The leading term that violates the global $U(1)$ involving scalar fields is the last term of Eq. (\ref{eq:pot1model3}). Owing to its high dimensionality its contribution to the shift in $\overline{\theta}$ is small, which is why the model has no axion quality problem. This issue is discussed in more detail in Sec.~\ref{sec:quality3}.

The higher dimensional operators with inverse Planck mass suprpession shown in Eqs. (\ref{eq:Yuk}) and (\ref{eq:pot1model3}) are not the only ones allowed by the gauge symmetries.  Those shown in Eq. (\ref{eq:Yuk}) are the lowest order operators that generate masses for the $SO(10)$ singlet fermions $(N_a, \chi)$, while the last term of Eq. (\ref{eq:pot1model3}) involves only the scalars acquiring high scale VEVs of order $f_a$. These are the terms that can potentially upset the axion quality the most. Terms such as $(H^2)^3 S/M_{\rm Pl}^3$ and $(\overline{\Delta})^6 S/M_{\rm Pl}^3$ are also allowed. If we insert in the first term the electroweak VEV, the correction to the axion mass from here is of order $m_a \sim v^3/(f_a M_{\rm Pl}^3)^{1/2} \sim 10^{-18}$ eV, which is sufficiently small compared to the QCD contribution to keep $\overline{\theta} \leq 10^{-10}$. This can be seen from Eq. (\ref{eq:gravmass}), which yiels $\overline{\theta} \leq 10^{-26}$ from this term, where $m_a \simeq 10^{-5}$ eV is used. This term does induce through loop diagrams an effective $T^{12} S^*$ operator with only  three powers of $M_{\rm Pl}$ suppression (compared to nine powers in the last term of Eq. (\ref{eq:pot1model3})), which will be analyzed in Sec. \ref{Sec:loop}.

\subsection{Identifying the axion field}

The axion field should be orthogonal to the Goldstone bosons eaten up by the gauge bosons, as well as to physical scalar fields that acquire masses through the Higgs potential.  In order to identify the axion filed, it is convenient to utilize the branching rules for $SO(10) \supset SU(5) \times U(1)_X$, which for the fields of interest are given by
\begin{eqnarray}
&~&10 \rightarrow 5(2) + \overline{5} (-2) \nonumber \\
&~&\overline{126} \rightarrow 1(10) + 5(2) + \overline{45} (-2) + ... \nonumber\\
&~&45 \rightarrow 1(0) + 24 (0) + ...
\end{eqnarray}
Here we have only kept fields that contain SM singlets or doublets, since only these components have neutral members that can acquire vacuum expectation values and thus be constituents of the axion field. We denote the SM doublet fields from the $H$ and $\overline{\Delta}$ and the SM singlets from $\overline{\Delta}$ and $A$ fields as:
\begin{eqnarray}
H &\supset& H_u^{10}(2,-2) + H_d^{10}(-2,-2) \nonumber \\
\overline{\Delta} &\supset& H_u^{126}((2,-2) + H_d^{126}(-2,-2) + 1_{126}(10,-2) \nonumber \\
 H' &\supset& H^{10'}(2,0) + H^{*{10'}}(-2,0) \nonumber \\
 A &\supset& 1_{45}(0,0) + 1'_{45}(0.0)~.
\end{eqnarray}
Here the $U(1)_X \times U(1)_a$ charges are indicated for each component. The $H^{10'}(2,0)$ is defined to have $Y/2 = +1/2$ and $H^{* 10'}(-2,0)$ is its conjugate.

It is straightforward to identify the combination of SM doublet and singlet fields that appear in the potential of Eq.~(\ref{eq:pot1model3}). In particular, the following terms involving the SM doublet fields arise when Eq.~(\ref{eq:pot1model3}) is expanded and the SM singlet VEVs are inserted:
\begin{eqnarray}
V &\supset& a_u\,( H_u^{126^\dagger} H_u^{10})
+ a_d\, (H_d^{126^\dagger} H_d^{10}) + a_{ud}\, (H_u^{126^\dagger} H_u^{10})(H_d^{126^\dagger} H_d^{10}) \nonumber\\
&+& a_{du}\,(H_u^{126^\dagger}H_u^{10})(H_d^{10^\dagger} H_d^{126}) + a_{uu}\,(H_u^{126^\dagger} H_u^{10})^2 \nonumber \\
&+& a_{dd}\,(H_d^{126^\dagger} H_d^{10})^2 + a_u'\,(H^{10'^\dagger} H_u^{10})\, T^2 + a_u'\,(H_d^{10} H^{10'}) \,T^2 + h.c.
\label{eq:pot2model3}
\end{eqnarray}
Each of the neutral field which acquires a VEV that is in general complex, can be parameterized as (in this section we denote the various VEVs as $v_i$ for consistency with the notation used in the GUT literature, rather than $f_i$ that was used for models I and II):
\begin{equation}
\Phi_i =\frac{1}{\sqrt{2}}(\rho_i + v_i e^{i\alpha_i}) e^{i \eta_i/v_i}
\end{equation}
where $v_i$ and $\eta_i$ are real.  Freezing the radial components $\rho_i$, which are irrelevant for axion identification, we parameterize
\begin{eqnarray}
&~&H_u^{10} = \frac{v_u e^{i \alpha_u}} {\sqrt{2}} e^{i \frac{\eta_u}{v_u}},~~~H_u^{\overline{126}} = \frac{v_u' e^{i \alpha_u'}} {\sqrt{2}} e^{i \frac{\eta_u'}{v_u'}},~~~H_d^{10} = \frac{v_d e^{i \alpha_d}} {\sqrt{2}} e^{i \frac{\eta_d}{v_d}},~~~H_d^{\overline{126}} = \frac{v_d' e^{i \alpha_d'}} {\sqrt{2}} e^{i \frac{\eta_d'}{v_d'}}, \nonumber \\
&~&\tilde{H}^{10} = \frac{\tilde{v} e^{i \tilde{\alpha}}} {\sqrt{2}} e^{i \frac{\tilde{\eta}}{\tilde{v}}},~~~\Delta_R^{\overline{126}} = \frac{v_R e^{i \alpha_R}} {\sqrt{2}} e^{i \frac{\eta_R}{v_R}},~~~S = \frac{v_S e^{i \alpha_S}} {\sqrt{2}} e^{i \frac{\eta_S}{v_S}},~~~T = \frac{v_T e^{i \alpha_T}} {\sqrt{2}} e^{i \frac{\eta_T}{v_T}}~.
\label{eq:parammodel3}
\end{eqnarray}
The two SM singlets from ${\bf 45}$ do not contribute to the axion field, and are left out here. By separate gauge and PQ rotations we can make the VEVs $(v_S,\,v_T,\,v_R)$ real (i.e., $\alpha_{S,T,R} = 0$ can be set), while the Higgs doublet VEVs are in general complex.

Inserting Eq.~(\ref{eq:parammodel3}) into Eq.~(\ref{eq:pot2model3}), we identify four massive pseudoscalar fields:
\begin{eqnarray}
{\cal A}_1&=&N_1\left[\frac{\eta_u}{v_u}-\frac{\eta'_u}{v'_u} \right] \nonumber \\
{\cal A}_2&=&N_2\left[\frac{\eta_d}{v_d}-\frac{\eta'_d}{v'_d} \right] \nonumber \\
{\cal A}_3&=&N_3\left[2\frac{\eta_T}{v_T} + \frac{\eta_u}{v_u} - \frac{\tilde{\eta}}{\tilde{v}}\right]\nonumber\\
{\cal A}_4&=&N_4\left[2\frac{\eta_T}{v_T} + \frac{\eta_d}{v_d} + \frac{\tilde{\eta}}{\tilde{v}}\right]~.
\label{eq:pseudo}
\end{eqnarray}
Here $N_{1-4}$ are normalization constants.
The three Goldstone fields, which are eaten up by the longitudinal components of $(Z,\, X,\,V_a)$ with $V_a$ being the gauge boson associated with $U(1)_a$, are identified as:
\begin{eqnarray}
    G_Z &=&N_Z(v_u \,\eta_u+v'_u\,\eta'_u-v_d \,\eta_d -v'_d \,\eta'_d + \tilde{v} \,\tilde{\eta})\nonumber\\
    G_X&=&N_X(10\, v_R\,\eta_R+2\,v_u\, \eta_u+2\,v'_u\, \eta'_u-2\,v_d\, \eta_d -2\,v'_d\,\eta'_d + 2\, \tilde{v}\, \tilde{\eta})\nonumber\\
    V_a&=&N_a(-2\,v_R\, \eta_R +v_T \,\eta_T+12\,v_S \,\eta_S-2\,v_u\, \eta_u-2\,v'_u\, \eta'_u-2\,v_d\, \eta_d-2\,v'_d\, \eta'_d)~.
    \label{eq:Gold}
\end{eqnarray}
The axion field is the combination orthogonal to the seven fields listed in Eq. (\ref{eq:pseudo}) and Eq. (\ref{eq:Gold}), which is obtained to be:
\begin{eqnarray}
a = N( c_S \eta_S + c_T \eta_T+ c_u \eta_u +c_u' \eta_u' + c_d \eta_d + c_d' \eta_d' + \tilde{c} \tilde{\eta}).~
\label{eq:afield}
\end{eqnarray}
It is convenient to make a few definitions in order to write down the coefficients $c_\alpha$ appearing in Eq. (\ref{eq:afield}):
\begin{eqnarray}
&& V_u^2 =v _u^2 + v_u'^2,~~V_d^2 = v_d^2 + v_d'^2,~~v^2 = V_u^2+V_d^2+\tilde{v}^2 \nonumber \\
&& X = v_T^2 \,v^2 + 4\,\tilde{v}^2(V_u^2+V_d^2) + 16 \,V_u^2 V_d^2 \nonumber \\
&&\tan\beta_u = \frac{v_u'}{v_u}, ~\tan\beta_d = \frac{v_d'}{v_d}
\end{eqnarray}
with $v^2 = (174~{\rm GeV})^2$. In terms of these quantities, the coefficients $c_\alpha$ and the normalization constant $N$ in Eq. (\ref{eq:afield}) can be written as:
\begin{eqnarray}
&&c_S = X,~~ c_T = -12\, v_S\, v_T \,v^2 \nonumber \\
&&c_u = 24\, v_S V_u(2\, V_d^2 + \tilde{v}^2) \cos\beta_u \nonumber \\
&&c_u' = 24 \,v_S V_u(2\, V_d^2+ \tilde{v}^2) \sin\beta_u\nonumber \\
&&c_d = 24 \,v_S V_d(2 \,V_u^2 + \tilde{v}^2) \cos\beta_d \nonumber \\
&&c_d' = 24 \,v_S V_d (2\,V_u^2+\tilde{v}^2) \sin\beta_d \nonumber\\
&&\tilde{c} = 24\, v_S \tilde{v} (V_d^2-V_u^2)\nonumber\\
&&N = 1/\sqrt{X(X+ 144 v_S^2 v^2)}~.
\label{eq:Cimodel3}
\end{eqnarray}
It is interesting to note that the $\eta_R$ field has  disappeared from $a$. Furthermore, since $v^2 \ll v_S^2, v_T^2$ we have $X \simeq v_T^2 \,v^2$ and $N \simeq 1/(v_T^2 \,v^2 \sqrt{1+ 144 v_S^2/v_T^2})$.

It is straightforward to compute the coupling of the axion to the gluon fields.  This coupling is induced through the $\eta_{u,d}$ and $\eta_{u',d'}$ content of Eq. (\ref{eq:afield}) through their couplings to the quarks in {\bf 16}-fermions as well as from the $\eta_S$ component through its couplings to the quarks in the {\bf 10}-fermion. Our calculation shows
\begin{eqnarray}
{\cal L}_{a G\tilde{G}} = \frac{g_s^2 }{32\pi^2} \frac{a}{f_a}\, G_{\mu \nu}^a\tilde{G}^{a, \mu \nu},~~
f_a= v_S /\sqrt{1+\frac{144 v^2_Sv^2}{X}}.~~~
\label{eq:fa}
\end{eqnarray}
Note that the rotation matrices that are used to diagonalize the quark mass matrices do not appear in the axion coupling to the physical quarks in the model, since the coupling matrices are proportional to the quark mass matrices. 


 \section{Phenomenology of the SO(10) axion} 
 
\label{sec:Model3-pheno}

 In this section, we present some of the phenomenological implications of the $SO(10) \times U(1)_a$ model, starting with a discussion of the quality constraint on the scales of the model. We shall also see that the model interpolates between the KSVZ- and the DSFZ-axion models.  

 \subsection{Quality constraint}
\label{sec:quality3}

From the gauge quantum numbers of the various fields in the model (Cf: Table \ref{tab:tab3}), we find that the leading gravity-induced PQ symmetry breaking operator involving scalar fields that acquire large VEVs is
\begin{equation}
V_{\rm gravity} = \frac{\kappa\, e^{i \delta}\, T^{12}\, S^*}{(12)!\, M_{\rm Pl}^9} + h.c.
\label{eq:grav}
\end{equation}
This shifts the minimum of the axion potential from $\overline{\theta}=0$ to a finite value given by
\begin{equation}
\overline{\theta} \simeq \frac{\kappa \sin\delta}{(12)!\, 2^{11/2}} \frac{f_a\, v_T^{12}}{M_{\rm Pl}^9\, m_\pi^2\, f_\pi^2} \frac{(m_u+m_d)^2}{m_u\,m_d}
\frac{\left(1+ \frac{144 \,v_S^2}{v_T^2} \right)} {\sqrt{1+ \frac{144\, v_S^2 \,v^2}{X}}}~.
\end{equation}
Defining 
\begin{eqnarray}
    r\equiv \frac{v_S^2}{v_T^2},
    \end{eqnarray}
and using the approximate relation $X \simeq v_T^2 v^2$, this simplifies to
\begin{eqnarray}
\overline{\theta} \simeq \left(\frac{\kappa \sin\delta}{(12)!\,2^{11/2}}\right) \left(\frac{f_a^{13}}{f_\pi^2\, m_\pi^2\, M_{\rm Pl}^9}\right) \frac{(m_u+m_d)^2} {m_u\, m_d} \left(\frac{(1+144 r)^{13/2}}{r^6}\right)
\label{eq:theta10}
\end{eqnarray}
The last factor which depends on the ratio $r$ in this equation has a minimum at $r = 1/12$, in which case its value is $14412774445056 \sqrt{13}=5.1966 \times 10^{13}$. Using $\kappa \sin\delta = 1$, $M_{\rm Pl} = 1.22 \times 10^{19}$ GeV, the maximum value for $f_a$ that keeps $\overline{\theta} \leq 10^{-10}$ is found to be $f_a < 6.96 \times 10^{11}$ GeV, where the minimum value of the function of $r$ is assumed. If the symmetry factor of $1/12!$ is not included in Eq. (\ref{eq:grav}), the largest allowed value of $f_a$ would be reduced by about a factor of 4, to $f_a \leq 1.5  \times 10^{11}$ GeV. We see that the axion quality remains high from the leading scalar operator.

To address what range of the ratio $r$ is compatible with axion quality, we note that if $r\ll 1$, axion is mostly in the $S$ field and we have $f_a \simeq v_S$.  In the opposite limit, where $r \gg 1$, the axion lies mainly in the field $T$ and $f_a \simeq 12 v_T$.  For $r \ll 1$, we find from Eq. (\ref{eq:theta10}) that there is a lower limit of $r \geq 1.7 \times 10^{-5}$, where $f_a = 5 \times 10^{10}$ GeV is used, consistent with axion being the dark matter of the universe.  This corresponds to $v_S/v_T \geq 0.004$.  In the absence of $12!$ in the denominator of Eq. (\ref{eq:theta10}), $v_S/v_T \geq 0.02$ will be required.  As we shall explain in more detail in Sect.~\ref{sec:hybrid}, in the limit $r\rightarrow 0$ the model would be reduced to the KSVZ axion model.  But axion quality does not allow the model to approach this limit. In the case where $r \gg 1$, there is only a mild constraint that $v_S \leq M_{\rm Pl}$, which would allow $r$ to be as large as $10^9$, consistent with axion quality.


\subsection{Loop effects on quality constraint} 
\label{Sec:loop}

\begin{figure}[t!]
		\centering
		\includegraphics[width=0.8\textwidth]{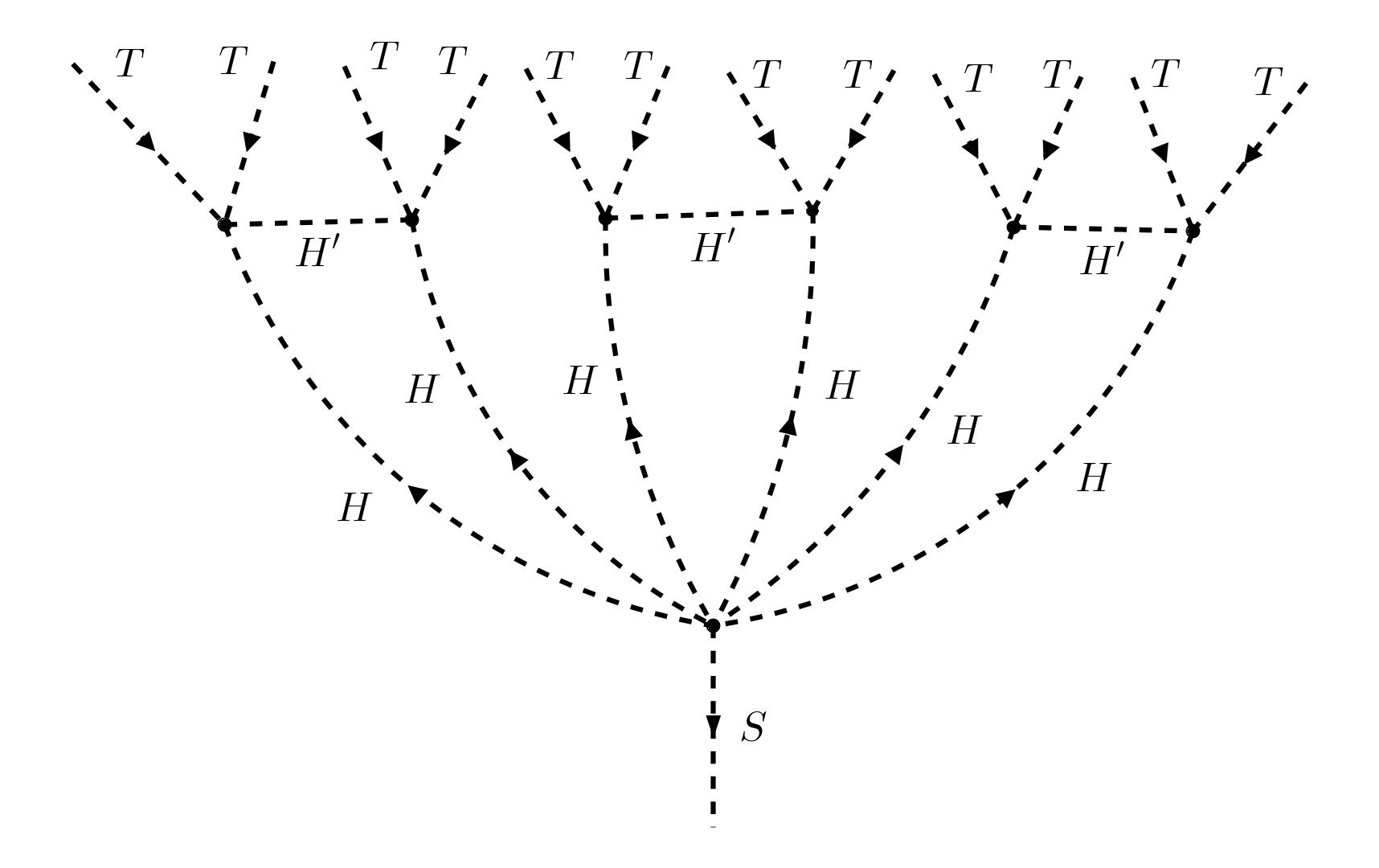}
	\caption{Three loop diagram that contributes to $U(1)_{\rm PQ}$ symmetry breaking in the $SO(10)$ model.} 
	\label{fig:threeloopmodel3}
\end{figure}

Let us now turn to the loop effects on the quality constraint.
In our model, an operator of the form $H^6 S/M^3_{\rm Pl}$ is allowed by the gauge symmetry. While it does not affect the axion quality at tree-level, when combined with the renormalizable coupling $(\lambda/2)\, H H' T^2$ of Eq. (\ref{eq:pot1model3}), this coupling would induce an operator $T^{12} S^*$ through a three-loop diagram shown in Fig.~\ref{fig:threeloopmodel3}. This induced operator has only three powers of $M_{\rm Pl}$ suppression, as opposed to the nine powers seen in Eq.~(\ref{eq:grav}).
Writing the $H H' T^2$ coupling as $\lambda/2(H_u H_d + H_c H_c^*) T^2$, with $H_c$ denoting the color triplet scalar, we find that the three-loop induced correction to the scalar potential that breaks the PQ symmetry, taking advantage of the $d=7$ Planck-induced operator $\gamma H^6 S/(6! M_{\rm Pl}^3)$, to be
\begin{eqnarray}
V \supset g_n \gamma \left(\frac{\lambda^2}{16\pi^2}\right)^3 \frac{T^{12} S^*}{12!\, M^6_{H'}M_{\rm Pl}^3}\left\{\frac{3}{2} + 2 \,{\rm ln} \left(\frac{M_{H_d}^2 - 4 M_T^2} {M_{H'}^2} \right)\right\}^3~.
\label{eq:loop}
\end{eqnarray}
Here the first term in the curly bracket arises from the exchange of color triplets, which are assumed to have GUT scale mass, taken to be the same as the $H'$ mass.  The second term arises from the exchange of $SU(2)_L$ doublets $H_u$ and  $H_d$, with $H_u$ mass being close to the weak scale and thus ignored.  (Note that the light Higgs doublet is almost entirely arising from $H_u$ in this model.) The mass of $H_d$ from $10_H$ is of the same order as the mass of $T$, and we have assumed it in the derivation of Eq. (\ref{eq:loop}) that $M_{H'} \gg M_{H_d} > 2 M_T$. The symmetry factor $g_n$ in Eq. (\ref{eq:loop}) is found to be
\begin{equation}
g_n = \frac{1}{3!} \frac{1.3.5}{(2!)^3} 6! =225.
\end{equation}
Choosing as an example $M_{H'} = 2 \times 10^{16}$ GeV, $M_T = 1 \times 10^{11}$ GeV and $M_{H_d} = 6 \times 10^{11}$ GeV, we find that these loop-induced contributions become comparable to the $d=12$ gravity-induced contribution of Eq. (\ref{eq:grav}) to $\overline{\theta}$ when the quartic coupling $\lambda \geq 0.003$ (for $|\gamma| \simeq |\kappa|=1$).  Thus, these loop corrections, while important, do not alter the quality of the axion. 

There is also another Planck suppressed operator with only three powers of inverse $M_{\rm Pl}$ of the form $(\Delta)^6 S^*/M^3_{\rm Pl}$, but this operator does not lead to anything that would affect the quality constraint like the operator $H^6S/M_{\rm Pl}^3$.

\begin{figure}[t!]
		\centering
\includegraphics[width=0.8\textwidth]{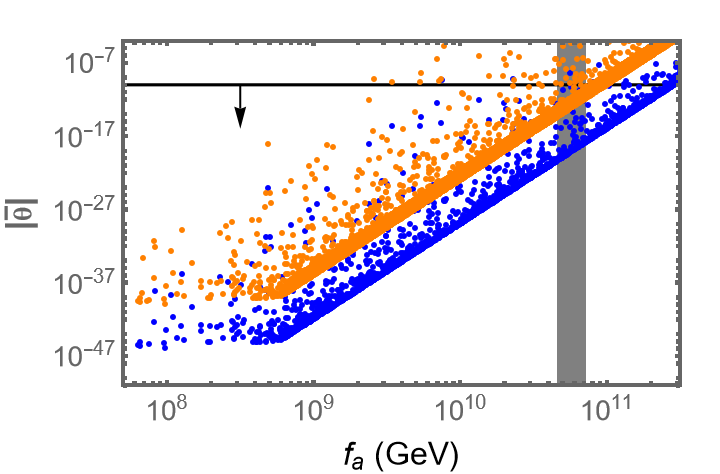}
	\caption{$|\bar{\theta}|$ vs $f_a$. $\bar{\theta}$ includes the three-loop contribution in Fig.~\ref{fig:threeloopmodel3}. The orange and blue points are shown for $\lambda=0.1$ and 0.01 respectively. The $|\theta|\leq 10^{-10}$ region is indicated by the arrow. The gray shaded band denotes the DM abundance constraint (Cf: Eq. (\ref{eq:fa1})).} 
	\label{Fig:theta}
\end{figure}

\begin{figure}[t!]
		\centering
		\includegraphics[width=0.8\textwidth]{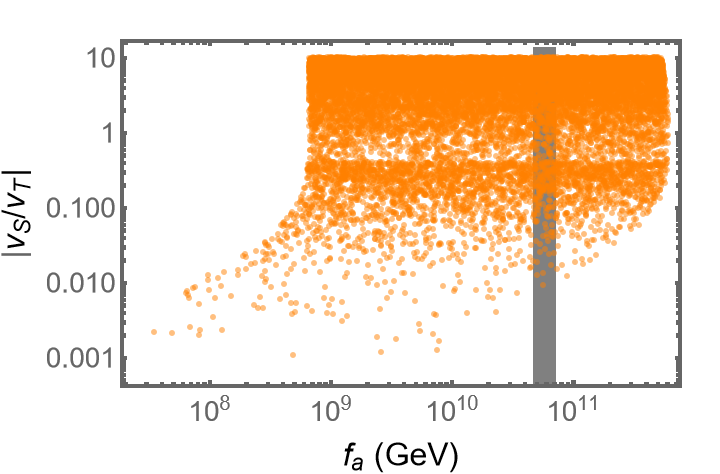}
	\caption{In this figure, we have imposed the quality constraint on the parameters of the model and present the allowed values of the $f_a$ as a function of the ratio of the two high scale VEVs $v_S$ and $v_T$ while varying other parameters as described in the text.  The gray shaded band denotes the values of $f_a$ for which the DM abundance (Cf: Eq. (\ref{eq:fa1})) is satisfied.}
	\label{Fig:vsvt}
\end{figure} 

\subsection{Phenomenological implications of the \boldmath{$SO(10)$} axion model}

In this section, we explore the parameter space of the $SO(10)$ model that is compatible with axion quality.  We first identify the couplings of the axion to the photon, electron and the nucleon, and then numerically study their correlations.

\subsubsection{Axion couplings to photon and fermions in the \boldmath{$SO(10)$} model}

To study phenomenological implications of the $SO(10) \times U(1)_a$ axion model, we first calculate the couplings of the axion to two photons as well as to fermions ($e,p,n$).  The general interaction of the axion with these fields is given in Eqs. (\ref{eq:acoup})-(\ref{eq:acoup3}) with the coupling coefficients defined in Eqs. (\ref{eq:acoup2})-(\ref{eq:acoup4}). The axion-photon coupling $C_{a\gamma}$ in the model defined in Eq. (\ref{eq:acoup2}) takes the value $C_{a\gamma} = 8/3-1.92$, where the second model-independent term comes from the non-perturbative QCD effects such as $a-\pi^0$ mixing~\cite{Srednicki:1985xd,Georgi:1986df,GrillidiCortona:2015jxo,DiLuzio:2021pxd}. It may appear surprising to see that $E/N=8/3$ in this model, which is the same value as in the DFSZ model, even in presnece of an additional {\bf 10} fermion.  This is a generic feature of GUT models which all give $E/N = 8/3$, as long as the embedding of electromagnetism is as in $SU(5)$, $SO(10)$ or $E_6$ models, yielding $\sin^2\theta_W = 3/8$ at the GUT scale~\cite{Srednicki:1985xd,Agrawal:2022lsp,DiLuzio:2024xnt}.  In the $SO(10)$ model it arises since the E/N ratio from the {\bf 16} fermions is the same as for the {\bf 10} fermions, each equal to 8/3, which then becomes an overall factor compared to the gluonic coupling.

The axion coupling to electron and nucleon have the form given in  Eq.~(\ref{eq:acoup2}) with the $C_{af}$-factors given by
\begin{eqnarray}
C_{ae} &=& \frac{24 r}{1+144 r} K_e + \frac{3 \alpha^2}{4\pi^2} \left[\frac{E}{N}\,{\rm log}\left(\frac{f_a}{m_e}\right) -1.92\, {\rm log}\left(\frac{\rm GeV}{m_e}\right)\right] \nonumber\\
C_{ap} &=& -0.47 + \frac{r}{1+144 r}(20.75 K_u -10.49 K_e) \nonumber \\
C_{an} &=& -0.02 + \frac{r}{1+144 r}(19.99 K_e - 9.73 K_u) ~.
\label{eq:coupling}
\end{eqnarray}
Here we have defined
\begin{equation}
K_u = \frac{2 {V_d}^2 + \tilde{v}^2}{v^2},~~~
K_e =  \frac{2{ V_u}^2 + \tilde{v}^2}{v^2}.
\label{kappa}
\end{equation}
$K_e$ has a range $(1.5-2)$ corresponding to $\tilde{v} = V_u$ and $\tilde{v} \ll V_u$. The value of $K_u$ can be much smaller, with an upper limit of 0.5 (corresponding to $\tilde{v} = v_u$). This gives a range $C_{ae} = (0.25-0.33)$ for $r=(0.1-1)$.

While the prediction of $g_{a\gamma}$ in our model is the same as in the DFSZ model, that for $g_{a e}$ is very different. The magnitude of the latter is much smaller, e.g., $g_{ae}$ has an upper limit arising from  $C^{max}_{ae} =0.33$.
If the electron coupling of the axion is measured to be larger than this value, our model will be ruled out. For comparison, note that in the KSVZ model $g_{ae} \approx 0$ and in the DFSZ-I model it is $(1/3){\rm sin}^2\beta(m_e/f_a)$ while in DFSZ-II it is $(-1/3) {\rm cos}^2\beta (m_e/f_a)$~\cite{DiLuzio:2020wdo}, where ${\rm tan}\beta= v_u/v_d$. (The DFSZ- I model is defined as one in which $H_d$ couples to leptons, whereas the DFSZ-II model has $\tilde{H}_u$ coupling to leptons.)
\begin{figure}[t!]
		\centering
		\includegraphics[width=0.8\textwidth]{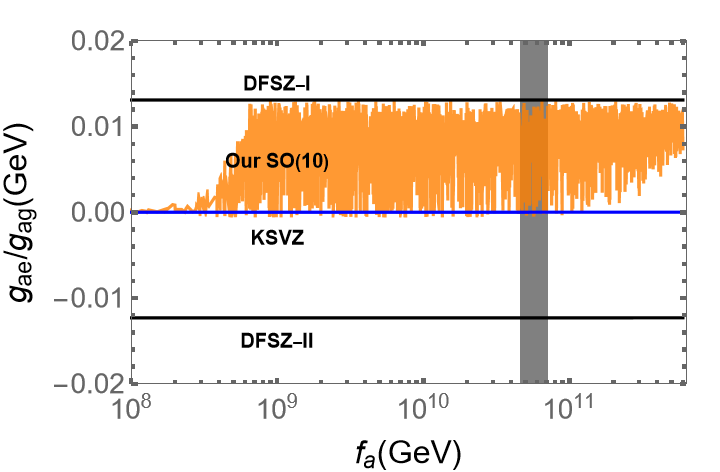}
	\caption{The plot of the ratio $g_{ae}/g_{ag}$ as a function of $f_a$ for our model (orange points). The region enclosed by the black line shows the DFSZ model range for this ratio, while the blue line shows the KSVZ model. Here we have used $\alpha_s(2~{\rm GeV}) = 0.32$ and varied $\tan\beta$ in the range $\tan\beta = \{0.25, 170\}$. We also show the region allowed by the DM abundance as a vertical shaded band.}
	\label{Fig:3}
\end{figure} 

\begin{figure}[t!]
		\centering
		\includegraphics[width=0.8\textwidth]{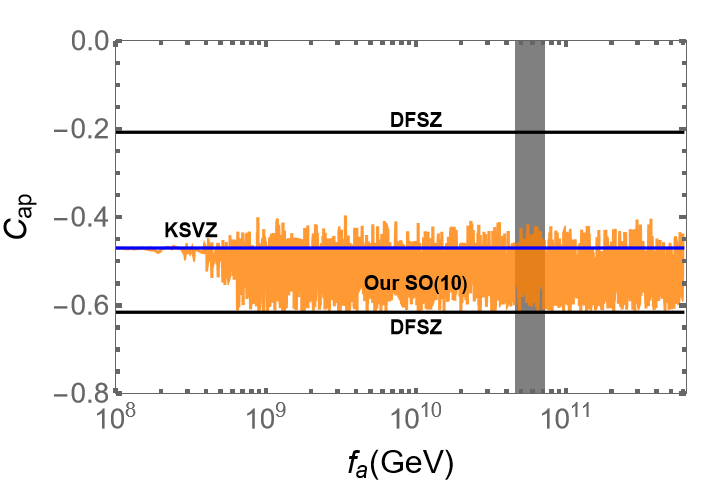}
	\caption{This plot shows the axion proton coupling as a function of the axion decay constant $f_a$. }
	\label{Fig:4}
\end{figure} 
\begin{figure}[t!]
		\centering
		\includegraphics[width=0.8\textwidth]{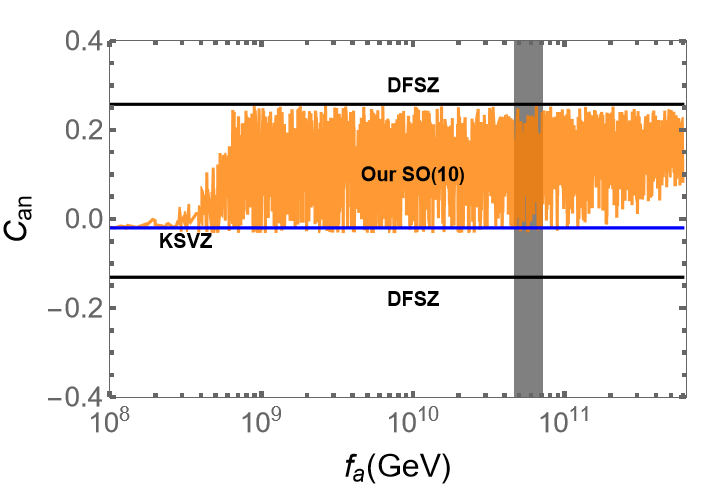}
	\caption{This plot gives the axion neutron coupling as a function of the axion decay constant $f_a$.}
	\label{Fig:5}
\end{figure}

\begin{figure}[t!]
		\centering		\includegraphics[width=0.8\textwidth]{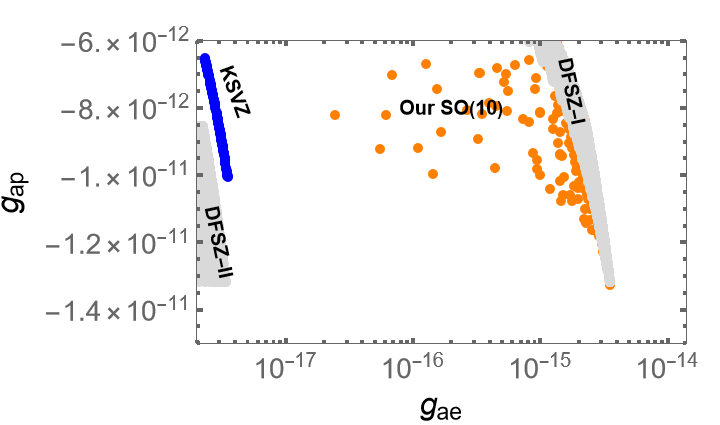}
\includegraphics[width=0.8\textwidth]{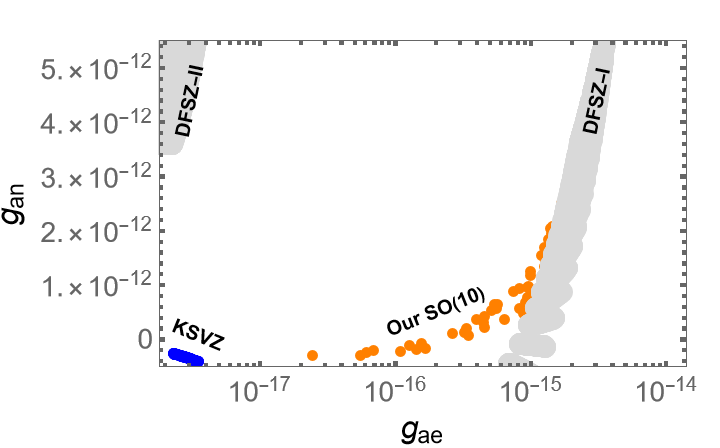}
	\caption{Axion proton (top)/neutron (bottom) vs axion-electron couplings for $4.6\times10^{10}\leq f_a (GeV)\leq 7.2\times 10^{10}$ where axions can be 100\% DM. }
	\label{Fig:6}
\end{figure} 
\vspace*{0.1in}

\subsubsection{Numerical results for the \boldmath{$SO(10)$} model parameter scan}

Here we present our results on a numerical scan of the model parameters to determine the couplings of the axion in the $SO(10)$ model. In order to derive these predictions, we vary all the VEVs, i.e., $v_S$, $v_T$, $V_{u,d}$ and $\tilde{v}$ while satisfying $v^2={V_u}^2+{V_d}^2+{\tilde{v}}^2$ and the angles ($\beta_u,\,\beta_d$). In our analysis we have used the constraints on the fermion mass fit which fixes the VEV ratios  $|v_u/v_d| \simeq 70.3$ and $v_u'/v_d' \simeq |18.1 + 3.7 i|$~\cite{Babu:2020tnf}. 

In Fig.~\ref{Fig:theta}, we present our results for $|\bar{\theta}|$  as a function of $f_a$ for two different values of $\lambda=0.1$ and 0.01 after including the loop contributions as given in Eq.~(\ref{eq:loop}). We also show here (gray band) the constraint of the relic abundance for $N_{\rm DW}=1$ as obtained in Ref.~\cite{Kawasaki:2014sqa,Ringwald:2015dsf}, i.e., $f_a=(4.6-7.2)\times 10^{10}$ GeV for $\Omega_a\rm{h^2}\simeq0.12$. This range of $f_a$ for axions providing 100\% DM is also consistent with the results of Ref.~\cite{Benabou:2024msj}. We find that the choice of $\lambda$ is dependent on $f_a$ and both the values of $\lambda$ are allowed by $|\bar{\theta}|\leq 10^{-10}$ and the relic abundance.

In Fig.~\ref{Fig:vsvt}, we plot the ratio $|v_S/v_T|$ as a function of $f_a$. 
All shaded regions satisfy the quality constraint.  As shown in Fig.~\ref{Fig:theta}, $\lambda$ can be chosen for each of the model points to have $|\bar{\theta}|\leq 10^{-10}$. The values of $|v_S/v_T|$ lie mostly in the range 0.1 to 10, although $v_S/v_T$ as small as 0.004 is also found, consistent with the discussion given in the last paragraph of Sec. \ref{sec:quality3}.  Also shown here in a gray band is the region consistent with axion dark matter abundance. 

We present the coupling ratio $\frac{g_{ae}}{g_{ag}}$ (where  $g_{ag}\equiv\frac{\alpha_s}{2\pi f_a}$ is the axion-gluon coupling) as a function $f_a$ in Fig.~\ref{Fig:3} and find that our model predicts a positive value for this quantity.
The predictions of the $SO(10)$ model are not allowed in the DFSZ-II-permitted regions.  
Similarly, $C_{ap}$ and $C_{ae}$ are plotted in Figs. \ref{Fig:4} and \ref{Fig:5}, respectively.  If the experimental values of these axion-nucleon couplings fall outside the orange-shaded regions of these figures, 
the $SO(10)$  model will be ruled out.

In Fig.~\ref{Fig:6} we show $g_{ae}$ vs axion-proton (neutron) coupling $g_{ap,n}\equiv m_{p,n} C_{ap,n}/f_a$ for $f_a$ satisfying the DM constraint. The correlated model points for this range of $f_a$ can be distinguished from KSVZ and DFSZ-II but overlap with the DFSZ-I model. This DM allowed region requires mainly $\lambda\sim 0.1$ to satisfy $|\bar{\theta}|\leq 10^{-10}$.

\subsection{SO(10) model as a Hybrid axion model}
\label{sec:hybrid}

In this subsection, we argue that our $SO(10)$ model is a hybrid between the KSVZ and DFSZ models. We note that the ratio $r=\frac{v^2_S}{v^2_T}$ interpolates between the KSVZ and DFSZ models, as can be seen from the axion composition of the model in Eq.~(\ref{eq:afield}). As $r\to 0$, the axion mainly consists of the field $\eta_S$ (case (i)), whereas in the other limit, when $r\to \infty$ the axion mainly consists of $\eta_T$ (case(ii)). Looking at the Yukawa couplings in Eq.~(\ref{eq:Yuk}) and Higgs potential in Eq.~(\ref{pot}), we see that in case (i), the singlet $S$ that contains the axion couples mostly to vector-like {\bf 10}-plet fermions like in the KSVZ model whereas in case (ii), the axion, which is mostly in $T$, does not couple to SM quarks like in the DFSZ model.

This interpolation is also reflected in Figs.~\ref{Fig:3}-\ref{Fig:6}, and Eq.~(\ref{eq:coupling}). As $r\to 0$, the couplings to $e, p$ and $n$ reduce to the KSVZ values 
whereas for $r\to \infty$ it goes to DFSZ-I values.
This shows that $r$ is indeed the interpolating parameter of the model and for arbitrary $r$, the predictions for the above observables are in the orange region of Figs.~\ref{Fig:3}-\ref{Fig:6} and represent the hybrid nature of our model interpolating between KSVZ and DFSZ-I cases. However, it should be noted that the ratio $r$ cannot be set to zero, since the axion quality requires a lower limit $v_S/v_T > 0.004$. Thus, although the model interpolates between the KSVZ and DFSZ models, it cannot be strictly identified as the KSVZ model consistent with axion quality.

\subsection{Domain Wall in the $SO(10)$ model }

To get the domain wall number in the $SO(10)$ model, we employ the same prescription as in the case of model I using Eq.~(\ref{DW}), viz., 
\begin{equation}
N_{\rm DW} = {\rm minimum ~ integer} \left\{\frac{1}{f_a} \sum_i n_i\, c_i\, f_i~,~~~n_i \in {\cal Z} \right\}~.
\label{eq:dw10}
\end{equation}
In our $SO(10)$ model, there are seven $c_i$'s which are listed in Eq. (\ref{eq:Cimodel3}), and therefore there are seven $n_i$'s that appear in Eq. (\ref{eq:dw10}). It is easy to see that for $n_{2,....7}=0$ and $n_1=1$ we get a solution to Eq.~(\ref{eq:dw10}). The number of domain walls in this model is one. Therefore, the model has no cosmological domain wall problem. The presence of one domain wall helps with creating axions, with the range quoted in Eq. (\ref{eq:fa1}) corresponding to the axion being the entire dark matter content of the universe.

\subsection{Other comments on the \boldmath{$SO(10)$} model} 

Here we make a few remarks on cosmological and other issues related to the $SO(10)$ axion model.

{\bf 1.} The  fermion fields $N_{1,2,3}$ of the model have masses that lie in the sub-eV range, since they arise from Planck suppressed operators. As a result, these fields can affect BBN. 
However, the  $N_{1,2,3}$ fields interact with the SM particles only by the exchange of the $U(1)_a$ gauge boson.  Since this gauge boson mass is of order  $v_a \sim 10^{12}$ GeV, the $N_i$'s go out of equilibrium at $T_*\simeq v_a(v_a/M_{\rm Pl})^{1/3}\simeq 10^9$ GeV. As a result, their contribution to the energy density at the epoch of BBN is small with $\Delta N_{eff}\simeq 0.13$. This prediction is in agreement with current CMB observations, but will be tested in the future in the  next generation CMB experiments.

{\bf 2.} The singlet $\chi$ in our model, with its mass of order TeV, can potentially overclose the universe. This is because its interactions with the SM paticles are very weak at $T \sim M_\chi$. One resolution to this problem is to give $\chi$ a large mass, of order the PQ breaking scale, by introducing an $SO(10)$ singlet fermion $N_0$ with zero $U(1)_a$ charge.
This new field would not affect the anomaly cancellation conditions, or other properties of the model.  A new Yukawa coupling of the type $\chi N_0 S^*$ would now be allowed, which would give the $\chi-N_0$ pair a Dirac mass of order $10^{12}$ GeV.  Additionally, a dimension-six four-fermion operator of the form $\chi N_1N_1N_1/M^2_{\rm Pl}$ is allowed in the Lagrangian. The decay $\chi \to 3N_1$ can now  proceed since $N_1$ is light. The width for this decay is given by $\Gamma_\chi\sim (m^5_\chi/M_{\rm Pl}^4)/192\pi^3 \sim 10^{-20}$ GeV. This decay occurs well before BBN, and thus leaves the success of BBN unaffected.

{\bf 3.}  We briefly comment on the gauge symmetry breaking, which is very similar to $SO(10)$ models discussed in detail in the literature with {\bf 10} + {\bf 126} Higgs fields (see for instance Ref.~\cite{Babu:2015bna}).
The {\bf 45} + {\bf 54} Higgs fields breaks the SO(10) down to $SU(3)_c\times SU(2)_L\times SU(2)_R\times U(1)_{B-L}$ symmetry, which is subsequently broken down to the standard model by the {\bf 126}-Higgs field. Here the ${\bf 54}$ field is optional, which would enable consistent symmetry breaking with the tree-level Higgs potential. In its absence the {\bf 45} can break $SO(10)$ down to the left-right symmetric group, but requires loop corrections~\cite{Bertolini:2009qj,Jarkovska:2021jvw}. Unification of gauge couplings in this setup has been studied in Re.~\cite{Babu:2016bmy,Babu:2015bna,Jarkovska:2021jvw}.  The $U(1)_a$ gauge symmetry is broken by the singlet fields $T, S$ which also break the accidental $PQ$ symmetry.

In the symmetry breaking chain that we have adopted, viz., $SO(10) \xrightarrow{M_U} SU(3)_c \times SU(2)_L \times SU(2)_R \times U(1)_{B-L} \xrightarrow{M_I} SU(3)_c \times SU(2)_L \times U(1)_Y \xrightarrow{M_W} SU(3)_c \times U(1)_{\rm em}$, the GUT scale turns out to be $M_U \simeq 2 \times 10^{16}$ GeV. The unified gauge coupling is   $\alpha_U \simeq 1/40$~\cite{Deshpande:1992au}. These values are obtained by assuming minimal fine-tuning and ignoring threshold effects. Note that the {\bf 10} fermion has a mass of the order of PQ scale, which would contribute to the gauge coupling evolution above this scale.  However, its impact on $\alpha_U$ is less than a percent, and the unification scale is essentially unaltered. 
The corresponding proton lifetime is about $4 \times 10^{36}$ yrs. (see for eg. Ref.~\cite{Babu:2024jdw}), to be compared with the Super-Kamiokande limit of $2.4 \times 10^{34}$ yrs~\cite{Super-Kamiokande:2020wjk}.  The inclusion of threshold corrections on the unification scale can modify this prediction somewhat.  Nevertheless, there is enough freedom in the model to be consistent with proton lifetime limits. The constraint on $f_a$ that arises from axion quality does not restrict the predictions for proton decay. 

{\bf 4.} The neutrino masses as well as other SM fermion masses in this model follow from type-I seesaw. Extensive analysis has been carried out in the non-supersymmetric version of similar $SO(10)$ models has been carried out in the literature~~\cite{Babu:1992ia,Bajc:2001fe,Fukuyama:2002ch,Bajc:2002iw,Goh:2003sy,Goh:2003hf,Babu:2005ia,Bertolini:2004eq,Bertolini:2005qb,Bertolini:2006pe,Bajc:2008dc,Joshipura:2011nn,Dueck:2013gca,Altarelli:2013aqa, Fukuyama:2015kra,Babu:2018tfi,Ohlsson:2019sja,Babu:2020tnf}, and they apply to our model.  Note that the $SO(10)$ singlet fermions $(N_a, \chi)$ do not mix with the $\nu_i$ and $\nu_i^c$ of the ${\bf 16}_i$ fields, since the model has no Higgs field belonging to the ${\bf 16}_H$ representation.  
We also expect the standard leptogenesis scenarios to work for our case to generate matter-anti-matter asymmetry of the universe. 

{\bf 5.}  For the choice of parameters in the $SO(10)$ model, where $f_a\sim (4-7) \times  10^{10}$ GeV, with $N_{\rm DW} = 1$, the right amount of axion DM density of the universe is realized.

\section{Summary}
\label{sec:summary}

We have constructed a class of new axion models where the $U(1)_{PQ}$ symmetry arises as an accidental global symmetry, solving the strong CP problem. A major advantage of these models is that they solve the axion quality problem owing to the presence of a gauged axial $U(1)_a$ symmetry. The $U(1)_a$ gauge symmetry eliminates dangerous low-order gravity-induced $U(1)_{PQ}$-violating operators. We have presented three examples of such models. The first class (called model I) extends the KSVZ invisible axion model with a gauged $U(1)_a$ symmetry acting on a set of vector-like quarks. Their axial $U(1)$ charges help in realizing a global $U(1)_{\rm PQ}$ symmetry accidentally with a QCD anomaly.  A minimum of four vector-like quarks are needed in this model in order to achieve high quality axion.  We have also shown that the quality of the axion is protected to a sufficient degree even after including loop-induced corrections to the scalar potential.  We have also shwon that the domain wall number in model I is one, which poses no cosmological problems. This occurs even with multiple vector-like quarks present in the model, which is a distinct prediction compared to the KSVZ model with multiple quarks.

Our second class of models (model II), is a generalization of model I to a family of models.  Here, we have shown how the introduction of $(m+1)$ quark fields where $m>3$ solves the axion quality problem.  As in the case of model I, the domain wall number is one here as well.  Model II has a stronger suppression of the gravity-induced operators that affect axion quality. We have also compared these models with the Barr-Seckel model. 

The third model is based on $SO(10) \times U(1)_a$, which uses fermions belonging to ${\bf 16}_a, {\bf 10}$ as well as singlet representations under $SO(10)$. The adopted $SO(10)$ model is the well-studied non-supersymmetric version that utilizes a ${\bf \overline{126}}$ and a complex ${\bf 10}$ of Higgs bosons to generate quark and lepton masses.  Such a setup is quite predictive in the neutrino sector, which has been widely studied in the literature.  This $SO(10)$ axion model has an interesting feature that it interpolates between the KSVZ and the DFSZ axion models, as the parameter $r$, which is a ratio of singlet scalat VEVs, is varied, consistent with axion quality.  We have also presented the predictions of the model for axion couplings to photon and the fermions.  We have explored the parameter space of the model that is consistent with the axion being the total dark matter content of the universe.  Since the couplings of the axion to the nucleon and the electron are distinct from the conventional DFSZ and KSVZ models, measurement of these couplings could test the model.  As in models I and II, the domain wall number is shown to be one and thus non-problematic in the $SO(10)$ model as well.

\section*{Acknowledgement} We thank Vasja Susič and Subir Sarkar for discussions. The work of KSB is supported by the U.S. Department of Energy under grant number DE-SC0016013 and that of BD by the DOE Grant de-sc0010813.

\appendix

\section{Appendix: Barr-Seckel model}
\label{sec:Appendix-A}

Here we give a brief summary of the Barr-Seckel model of high quality axion~\cite{Barr:1992qq}, which may be useful for a comparison of the models presented here.  We also comment on the domain wall number in this framework and a possible way of evading the cosmological problems when $N_{\rm DW}$ turns out to be larger than one.

The model introduces $(p+q$) quarks which transform as $(3,1,0)$ under SM gauge symmetry.  $p$ and $q$ are relatively prime integers, with $(p+q) \geq 10$ needed to solve the quality problem.  All left-handed quark fields carry zero charge under a gauged $U(1)$ symmetry while the right-handed quarks carry nonzero charges. The $U(1)$ charges of new VLQs are given as
\begin{equation}
\{(p+q) \times Q_L(0) + 
p \times Q_R(q) + q \times Q_R(-p)\}~.
\end{equation}
Here $(p+q)$, $p$ and $q$ are the multiplicities of the quark fields.  All mixed anomalies cancel with this setup.  One needs to ensure that the cubic $[U(1)]^3$ anomaly vanishes as well, which is achieved by introducing SM singlet fermions.

Two scalar fields, $S(-q)$ and $T(+p)$ are employed, which generate all quark masses via the Yukawa couplings
\begin{equation}
{\cal L}_{\rm Yuk} = \sum_{i=1}^p \overline{Q}_{Li} Q_{Ri} S + \sum_{i=p+1}^{p+q} \overline{Q}_{Li} Q_{Ri}T+ h.c.
\end{equation}
Denoting $S = (\rho_S+f_S)/\sqrt{2} e^{i \eta_S/f_S}$ and $T = (\rho_T + f_T)/\sqrt{2} e^{i \eta_T/f_T}$, 
the Goldstone boson eaten by the $U(1)$ gauge boson and the orthogonal axion field are given by
\begin{eqnarray}
G =\frac{ p f_T \eta_T - q f_S \eta_S} {\sqrt{p^2 f_T^2 + q^2 f_S^2}}, ~~~~a = \frac{ p f_T \eta_S - q f_S \eta_T} {\sqrt{p^2 f_T^2 + q^2 f_S^2}}~.
\end{eqnarray}
The axion decay constant $f_a$ is readily computed as
\begin{equation}
    f_a =\frac{f_T f_S} {\sqrt{p^2 f_T^2 + q^2 f_S^2}}~.
\end{equation}

Gravity-induced axion potential arises from terms such as
\begin{equation}
V_{\rm gravity} = \frac{\kappa \,e^{i \delta}\, S^p\, T^q}{p! \,q!\, M_{\rm Pl}^{p+q-4}} + h.c.
\end{equation}
This leads to an axion potential given as
\begin{equation}
V_{\rm gravity}^{(a)} = \frac{2\,\kappa \,f_S^p \,f_T^q}{2^{(p+q)/2} p!\,q!} \frac{\cos(\frac{a}{f_a}+\delta)}{M_{\rm Pl}^{p+q-4}}~.
 \end{equation}
 Minimizing this potential, along with the QCD-induced axion potential, yields
 \begin{equation}
     \overline{\theta} \simeq \frac{2 \kappa \sin\delta}{2^{(p+q)/2}\, p!\, q!}\frac{ f_S^p f_T^q}{f_\pi^2 m_\pi^2} \frac{(m_u+m_d)^2}{m_u\, m_d} \frac{1}{M_{\rm Pl}^{p+q-4}}~.
 \end{equation}
 If we choose $f_S = f_T = f_a \sqrt{p^2+q^2}$, the following values of $\overline{\theta}$ are obtained corresponding to different choices of $(p,\,q)$ and $f_a$, all for $\kappa \sin\delta = 1$:
\begin{equation}
  |\overline{\theta}| \simeq
    \begin{cases}
      5 \times 10^{-18} ~~ (p=1,\,q=9,,\,f_a = 10^9~{\rm GeV})\\
      1 \times 10^{-17} ~~(p=3,\,q=7, \, f_a = 10^{9}~{\rm GeV})\\
      4 \times 10^{-15}~~~(p=1,\,q=13,\,f_a=10^{12}~{\rm GeV})~.
    \end{cases}       
\end{equation}
This results in a high quality axion.

Since $p+q = 10$ is the minimal choice of VLQs in the framework, we have investigated ways of UV-completing the model with the choice $p=1$ and $q=9$, which are co-prime. Three singlet fermions with $U(1)$ charges of $(1,\,15,\,-16)$ will cancel the linear anomaly, as well as the cubic anomaly.  In the case, the Higgs fields $S$ and $T$ have charges $1$ and $9$.  The couplings $N(15) N(-16) S$ and $N(1) N(1) S^2/M_{\rm Pl}$ would generate masses for these singlets.  

The domain wall number in this model can be found to be
\begin{equation}
N_{\rm DW} = {\rm minimum ~integer}(n_1 q+ n_2 p)~.
\end{equation}
For the case of $(q,p) = (1,9)$ one has $N_{\rm DW} = 1$, which can be verified by choosing $(n_1,n_2) = (1,0)$.  On the other hand, if $(q,p) = (3,7)$, which are also co-prime integers, $N_{\rm DW} = 3$.  We briefly outline how cosmological problem with $N_{\rm DW} = 3$ may be evaded by using higher dimensional operators that break $U(1)_{\rm PQ}$ explicitly.

\subsection{Fate of the Domain Walls where \boldmath{$N_{\rm DW}=3$}}

Since this model has a Planck induced PQ breaking term, the different vacua will not be degenerate leading to decay of the domain walls as the universe evolves. The strength of this breaking term is 
given, in the model with $(q,p)=(3,7)$,  by $\Delta V\simeq \frac{f_S^7 f_T^3}{M^6_{\rm Pl}}$. This leads to a difference in the potential between the degenerate vacua by this amount $\Delta V$. This energy difference leads to decay of the domain walls with a lifetime for the domain walls to be~\cite{Hiramatsu:2013qaa}
$t_{decay}\sim \frac{\sigma_W}{\Delta V}$. Here $\sigma_W$ is the energy per unit area in the wall and is given by $\sigma_W\sim 9 m_a f^2_a$~\cite{Hiramatsu:2012sc, Zhang:2023gfu}. We must require that $t_{decay} \leq t_{BBN}$ to maintain the success of BBN. Also in this model the domain walls dominate the energy density of the universe at $t_{DW}\simeq\frac{M^2_P}{8\pi \sigma_{DW}}$ and we must demand that $t_{decay} \leq t_{DW}$, so as to not  alter the course of the evolution of the universe.

Let us try to see if in this model, the constraint $t_{decay} \leq t_{BBN}$ and $t_{decay} \leq t_{DW}$  are satisfied. For this, we choose $f_S= f_T$ , taking $f_a\simeq 10^{10}$ GeV. We get $t_{decay}\sim 10^{-2}$ seconds. This therefore satisfies the constraint that domain walls do not affect BBN.   Similarly, for the same choice of parameters, we get $t_{DW}\simeq 10^3$ sec. satisfying the requirement $t_{decay}\leq t_{DW}$.\footnote{See however Ref.~\cite{Beyer:2022ywc} where it has been shown that the production of strings by decaying domain walls for the case $N_{DW}\geq 2$ leads to a bound of $f_a \leq 5\times 10^8$ GeV.}

 \bibliographystyle{style}
 \bibliography{BIB.bib}

 \end{document}